\documentclass{emulateapj}

\def\kms{\,km\,s$^{-1}$}
\def\mo{M$_\odot$\,}

\def\Dwa{$\,$\uppercase\expandafter{\romannumeral5}$\,$}

\def\sles{\lower2pt\hbox{$\buildrel {\scriptstyle <}
   \over {\scriptstyle\sim}$}}

\def\sgreat{\lower2pt\hbox{$\buildrel {\scriptstyle >}
   \over {\scriptstyle\sim}$}}
\def\sharpnull#1{}

\begin{document}

\title{Magnetically-driven explosions of rapidly-rotating white dwarfs 
following Accretion-Induced Collapse}

\author{L. Dessart\altaffilmark{1},
A. Burrows\altaffilmark{1},
E. Livne\altaffilmark{2},
and C.D. Ott\altaffilmark{1}
}
\altaffiltext{1}{Department of Astronomy and Steward Observatory,
                 The University of Arizona, Tucson, AZ \ 85721;
                 luc@as.arizona.edu,burrows@as.arizona.edu,cott@as.arizona.edu}
\altaffiltext{2}{Racah Institute of Physics, The Hebrew University,
Jerusalem, Israel; eli@frodo.fiz.huji.ac.il}

\begin{abstract}

   We present 2D multi-group flux-limited diffusion  magnetohydrodynamics (MHD)
simulations of the Accretion-Induced Collapse (AIC) of a rapidly-rotating 1.92\,\mo white dwarf,
started from a 2D rotational equilibrium configuration. We focus on determining
the dynamical role of MHD processes after the formation of a millisecond-period 
protoneutron star. We find that, compared with the previous study of Dessart et al. 
that did not include magnetic fields, magnetic stresses can lead to
a powerful explosion with an energy of a few Bethe, rather than a weak one of at most 0.1\,B, 
with an associated ejecta mass of $\sim$0.1\,\mo, instead of a few 0.001\,\mo.
The core is spun down by $\sim$30\% within 500\,ms after bounce, and the rotational
energy extracted from the core is channeled into magnetic energy that generates a 
strong magnetically-driven wind, rather than a weak neutrino-driven wind.
Baryon loading of the ejecta, while this wind prevails, precludes it from becoming relativistic.
This suggests that a $\gamma$-ray burst is not expected to emerge from such AICs
during the early protoneutron star phase, except in the unlikely event that the
massive white dwarf in this AIC context has sufficient mass to lead to black hole formation.
In addition, we predict both negligible $^{56}$Ni-production (that should 
result in an optically-dark, adiabatically-cooled explosion) and 
the ejection of 0.1\,\mo of material with an electron fraction of 0.1--0.2. 
Such pollution by neutron-rich nuclei will put strong constraints on the
possible rate of such AICs. Moreover, being free from ``fallback,'' such highly-magnetized millisecond-period 
protoneutron stars may, as they cool and contract, become magnetars, and the 
magnetically-driven winds may later transition to Poynting-flux dominated and relativistic winds, 
eventually detectable as $\gamma$-Ray Bursts (GRBs) at cosmological distances. 
The likely low event rate of AICs, however, will constrain them to be, at best, a small subset of 
GRB and/or magnetar progenitors.

\end{abstract}


\section{Introduction}

   Some white dwarfs, located in binary systems, are thought to lead to
thermonuclear runaways and Type Ia supernovae, a circumstance in which a 
Chandrasekhar mass carbon and oxygen core is incinerated, on a timescale of $\sim$1\,s, 
to intermediate-mass and iron-peak elements. 
In the single-degenerate scenario, the white dwarf accretes
material from a hydrogen or helium donor star. Aided by the stabilizing effect of
fast and differential rotation (Yoon et al. 2004), it reaches the $\sim$1.4\,\mo Chandrasekhar limit
and explosively burns carbon and oxygen under degenerate conditions.
However, for sufficiently high mass accretion rates, carbon and oxygen may burn 
under non-degenerate conditions to allow the formation of a high-density ONeMg core which, upon 
achieving the Chandrasekhar mass, collapses on a dynamical timescale of $\sim$100\,ms
(Nomoto \& Kondo 1991; Nomoto et al. 1995; Gutierrez et al. 1996; Gil-Pons \& Garc{\'{\i}}a-Berro 2001).
In the double-degenerate scenario, two white dwarfs in a short-period binary system eventually
coalesce to form a massive white dwarf that exceeds the $Y_{\rm e}$-corrected Chandrasekhar mass limit. 
Smooth-Particle-Hydrodynamics (SPH) simulations (Benz et al. 1990; Mochkovitch \& Livio 1989, 1990; 
Segretain et al. 1997; Guerrero et al. 2004; Yoon et al. 2007)
predict the formation of a complex object composed of a cold, slowly-rotating core, surrounded 
by a hot and fast-rotating envelope, and, further out, by some residual mass in a Keplerian disk. 
Depending on the temperature in the envelope, the mass accretion rate on the newly-formed white dwarf, 
and the mass of each white dwarf component, either Type Ia or Accretion Induced Collapse (AIC) are possible.
Overall, the occurrence rate of the AIC of white dwarfs is difficult to determine reliably,
although it seems unlikely that they occur more frequently than once per 20--50 Type Ia
event (Yungelson \& Livio 1998,2000; Madau et al. 1998; Fryer et al. 1999; 
Nelemans et al. 2001ab; Blanc et al. 2004; Mannucci et al. 2005; Belczynski et al. 2005; 
Greggio 2005; Scannapieco \& Bildsten 2005; Dessart et al. 2006a, hereafter D06).

   In this work, we focus on the subset of accreting white dwarfs that form ONeMg cores, 
which, due to their high central density and mass, collapse to form a neutron star. 
The formation path followed, either through
accretion from, or coalescence with, a companion, suggests a large angular momentum budget
for the resulting white dwarf. Such white dwarfs rotate differentially, with a specific 
angular momentum that increases outward, and can be dynamically stable well above the canonical 
Chandrasekhar mass of $\sim$1.4\,\mo, reaching theoretical masses as high as 4.1\,\mo 
(Ostriker \& Bodenheimer 1968; Hachisu 1986; Yoon \& Langer 2005). The precise distribution
of the angular momentum inside the star is not accurately known, in particular since 
magnetic torques may spin down the core and enforce solid-body rotation,
producing an even more slowly rotating core than presently predicted by current
SPH simulations ($\sim$20\,s period). In any case, the overlying envelope is fast-rotating and
the resulting progenitor is strongly aspherical (Guerrero et al. 2004; Yoon et al. 2007)
\footnote{Observationally-inferred rotation rates of white dwarfs are, however, generally
modest, a fact attributed to the copious angular momentum losses during their evolution,
even for isolated white dwarfs (Koester et al. 1998; Berger et al. 2005; Valyavin et al. 2005; 
Karl et al. 2005).
Magnetic, fast-rotating  white dwarfs in binary systems may also undergo considerable spin-down,
caused by magnetic torquing of the envelope (Ikhsanov 1999).
Merger events show, by contrast, a potential for the production of fast-rotating white dwarfs, 
with properties more germane to the present discussion (Ferrario \& Wickramasinghe 2005).
If fast rotation of either the core or the envelope were compromised, 
the outcome of the AIC would be more suitably described by the simulation of D06
for the 1.46\,\mo model, rather than that for the 1.92\,\mo model described here or in D06.}.

   The collapse of ONeMg cores has been investigated in the past both in the context
of white dwarfs (Baron et al. 1987b; Woosley \& Baron 1992; Fryer et al. 1999; D06) 
and in the context of 8-10\,\mo stars (i.e., low-mass massive stars; Hillebrandt et al. 1984; 
Mayle \& Wilson 1988; Baron et al. 1987a; Kitaura et al. 2006; Burrows et al. 2007c). 
In D06, we presented results of radiation hydrodynamics simulations of the AIC of white dwarfs that 
extended these past investigations simultaneously to include multi-dimensionality,
multi-group flux-limited diffusion for multi-flavor neutrino transport, and, perhaps most importantly,
rapid rotation. We focused on the 1.46\,\mo and the 1.92\,\mo models of Yoon \& Langer (2005)
and found for the latter that fast rotation led to numerous striking features, e.g., 
aspherical collapse, aspherical bounce, polar-confined explosions, polar-enhanced neutrino 
radiation fields, gravity-darkened neutron stars, oblateness of the neutron stars formed, 
and a residual Keplerian disk. In this, the differential rotation 
that results from core collapse sets the stage whereby, from initial seeds 
in the poloidal component, a magnetic field may exponentially grow on a timescale of a rotation period 
by the magnetorotational instability (MRI; Balbus \& Hawley 1991; Pessah \& Psaltis 2005,2006ab; 
Etienne et al. 2006), acting on the poloidal and on the toroidal fields.
If equipartition between the magnetic energy and the differential rotational 
energy is reached at saturation, very substantial fields may be generated for the fastest rotators
(Shibata et al. 2006). The 1.92\,\mo AIC white dwarf 
studied in D06, with an initial $T/|W|$ of 10\% ($W$ is the gravitational
energy of the initial equilibrium configuration, and $T$ is the initial rotational kinetic energy), 
represents a good candidate for MHD effects after core bounce. 
Unfortunately, it is not feasible, even with modern supercomputers,
both to follow the multi-dimensional evolution of a multi-group neutrino field for three species 
and to resolve finely enough the scales over which the magnetorotational instability
operates. To circumvent this difficulty, we start with fields that are strong enough so as to
reach soon after core bounce (after a few tens of rotation periods) the saturation values 
they would have reached had the MRI operated. Magnetic-field amplification then results 
exclusively from compression and winding.
This is the approach followed in Burrows et al. (2007b), where we presented results for 
magnetically-driven supernova explosions of massive star progenitors in the context of fast rotation. 

   The central motive for this work is to go beyond the investigations of D06
and to study the potential role of MHD effects, adopting initial fields that will be commensurate
at saturation with values expected from equipartition with the free energy of rotation in the
differentially-rotating surface layers of the rotating collapsed core. 
Thus, we investigate the MHD effects that may arise during the collapse phase and the 
few hundreds of milliseconds that directly follow the formation of the neutron star. 
Guided by our simulations, we comment on past and current suggestions that the AIC of white 
dwarfs are potential Gamma-Ray Burst (GRB) and/or magnetar progenitors, or are important 
production sites of r-process nuclei.

  In the next section, we discuss (briefly) the VULCAN/2D code, and the numerical setup, as well as 
the 1.92\,\mo AIC model we focus on. 
The details of the code can be found in Livne et al. (2004,2007), Dessart et al. (2006b),
and Burrows et al. (2007a), and the AIC initial model properties are given in D06.
In \S3, we describe the results of our simulations, and compare 
a reference model with weak initial fields to another model with strong initial fields.
We discuss our results in the broader astrophysical context in \S4. In \S4.1, we
summarize the main MHD effects we identify, and in \S4.2 we discuss the potential nucleosynthetic yields.
In \S4.3, we discuss potential electromagnetic displays from AICs, and whether they can be
$\gamma$-ray bursts, and in \S4.4 we discuss the ultimate fate of these rare objects. 
Finally, in \S5, we present our conclusions.

\section{Model and progenitor properties}

  The models presented in this work derive from the 1.92\,\mo model of D06 (see 
their \S2), originally produced in Yoon \& Langer (2005). The initial density, temperature, 
composition, and 
angular velocity distributions are coincident with those used in D06. To summarize, the central 
density is $\rho_{\rm c}$ = 5 $\times$ 10$^{10}$\,g\,cm$^{-3}$, the temperature is given by
$ T(r,z) = T_{\rm c} (\rho_{\rm c}/\rho(r,z))^{0.35}\,$ with $T_{\rm c} = 1.3$ $\times$ 10$^{10}$\,K 
($r$ and $z$ are the standard cylindrical coordinates), the initial electron fraction ($Y_{\rm e}$) 
is 0.5 throughout the grid, 
and the central angular velocity is 20.4\,rad\,s$^{-1}$ (with a period of 0.3\,s).
To minimize the dynamical influence of the medium surrounding the white dwarf, which needs 
to be initially defined in our Eulerian approach, we assign it properties that correspond 
to the low-density low-temperature edge of our equation of state (EOS, see also below) table, 
i.e., a density of 10$^3$\,g\,cm$^{-3}$, and a temperature of 4 $\times$ 10$^8$\,K. Furthermore, 
we also assign it a negligible angular velocity.
Initially, the polar (equatorial) radius is 660\,km (2350\,km), the total angular momentum is 
0.1092\,B\,$\cdot$\,s (1\,B$\equiv$10$^{51}$erg), the total rotational energy $T$ is 1.057\,B, and 
the gravitational energy $|W|$ is 12.69\,B. This model thus starts with a ratio $T/|W|$ of 0.083,
indicative of the large angular momentum budget in the progenitor white dwarf.
We refer the reader to \S2 and Fig.\,1 of D06 for a detailed description of this 
initial model, with 2D representations of the original distribution of the density, temperature, 
and angular velocity.

The magnetic field morphology we adopt is uniform inside a sphere of radius 600\,km (the poloidal
component has a $z$-orientation, i.e., is aligned with the axis of rotation), and dipolar
beyond. Note that the original model of Yoon \& Langer (2005) corresponds to an equilibrium
configuration that results from the balance between gravitational, rotational, and thermal energy,
and conveys no information on the magnetic-field morphology and magnitude.
One may argue that magnetic torques would enforce solid-body rotation in the initial
model we adopt from Yoon \& Langer (2005), making our approach inadequate.
We propose in this work that even such super-Chandrasekhar white dwarfs are unlikely
to lead to black hole formation, so the differential rotation of our initial
model does not compromise our conclusions in this respect.
Our main assumption is that such torques may alter the rotational profile, but
not reduce significantly the overall rotational
energy budget of the white dwarf, preserving its status as a fast rotator.
Furthermore, the initial differential nature of the rotation is not essential since
gravitational collapse will naturally lead to it.
For the binary merger scenario, magnetic torques would likely not have sufficient time to
sizeably torque the white dwarf before core collapse.
Little is known about the core magnetic fields of white dwarfs, and even less on that fraction
of them that may lead to AIC and rotate fast, nor do we know how the poloidal and 
toroidal components relate. Again, the key here is that we assume that starting from a 
given initial seed poloidal field, the magnetorotational instability will result in the 
exponential growth of the magnetic energy after bounce, and then saturate, after 
a few tens of rotation periods, at the field we obtain here. 
By that time, we expect that the system will have lost memory of 
the magnetic field distribution it started with. 
In practice, we perform simulations with an initial poloidal field of 10$^{11}$\,G for model B11 
and of 10$^{12}$\,G for model B12, while in both cases we use a toroidal field of 10$^{8}$\,G 
(these values are at the white dwarf center). 
Note that the magnetic field amplification at bounce resulting from compression alone
will lead to an enhancement of the magnetic field by a factor of only
$(\rho_{\rm nuclear}/\rho_{\rm c})^{(2/3)} \sim 300$, because of the very high initial
central density ($\rho_{\rm c} = 5 \times 10^{10}$\,g\,cm$^{-3}$). For a star with
an initial central density of 5 $\times 10^{9}$\,g\,cm$^{-3}$, a typical value at the onset of
core collapse for a massive-star progenitor, the field amplification would be by a factor of
$\sim$1200 instead, so that an initial field of 2 $\times 10^{11}$\,G would achieve
a comparable magnetic field at bounce. Hence, initial field values must always be interpreted
together with the initial central densities to properly gauging the amplification generated
by compression. The choice of a predominantly poloidal component is essential in our approach,
allowing for the growth of the toroidal component after the initial compression phase.
If the MRI operated, the exponential growth of both components of the magnetic field, permitted
by the large budget of free energy of rotation, would reach saturation values that are
orders of magnitude above the initial seed values. Hence, in this context, the main
limiting factor is the rotational energy budget rather than the initial properties
of the field.
The evolution of the B11 model is similar
to that of the non-magnetic 1.92\,\mo AIC model presented in D06, 
and we use it to gauge the importance of MHD effects in the B12 model.
The lack of magnetic effecs in the B11 model is a clear sign that
we do not resolve the MRI, since otherwise we would observe magnetic fields
that saturate at values similar to those seen in the B12 model, and we do not.
Note that the 1.92\,\mo progenitor model we discuss in this work is not meant to be 
representative of AICs in Nature, as, indeed, these may come with a range of
angular momenta or masses. But our B12 model is chosen because it is one example where all
conditions are prime for the generation of MHD effects, thus occupying one extreme
when compared to, for example, the more slowly rotating and quasi-spherical 1.46\,\mo model 
simulated in D06. This approach allows us, we hope, to bracket the range of outcomes for AICs.

  We adopt the same spatial grid as D06, with a Cartesian mapping of the 
region inside a spherical radius of 20\,km and a spherical-polar grid out to a 
maximum radius of 4000\,km. The maximum resolution of 0.48\,km is reached in the Cartesian inner region, 
and the minimum resolution is 100\,km at the maximum radius, with 71 regularly-spaced angular zones 
covering 90$^{\circ}$. In D06, we found that the top/bottom symmetry was perfectly 
preserved in the 180$^{\circ}$ simulations of the 1.46\,\mo AIC model. We speculate that the 
post-bounce time to explosion is so short that convection and the Standing-Accretion-Shock-Instability
(SASI; Blondin et al. 2003; Foglizzo et al. 2006)
cannot grow to any sizable level, perhaps inhibited by the strong differential rotation 
(Burrows et al. 2007b). Hence, we save time and opt for higher resolution 
by restricting our investigation to one hemisphere.

  The simulations we present here constitute the state-of-the-art, combining in the dynamical equations 
both the effects of magnetic field and neutrino transport. As in D06,
we employ the approximation of Multi-Group Flux-Limited Diffusion, using 16 logarithmically-spaced energy 
groups between 2.5\,MeV and 220\,MeV, for three neutrino species 
(electron neutrino $\nu_{\rm e}$, anti-electron neutrino $\bar{\nu}_{\rm e}$, and a third 
entity grouping the remaining neutrinos, designated ``$\nu_{\mu}$''). 
We also employ a tabulated version of the Shen (1998) EOS (see Dessart et al. 2006b for details). 
Finally, the gravitational potential is computed using a multipole solver, with a decomposition
in spherical harmonics that extends to $l=32$. Again, the VULCAN/2D code and our standard approach to
modeling core collapse are described in Livne et al. (2004,2007), Dessart et al. (2006b), and 
Burrows et al. (2007a), while further details on the AIC of white dwarfs have been given 
in D06.

\begin{figure*}
\plottwo{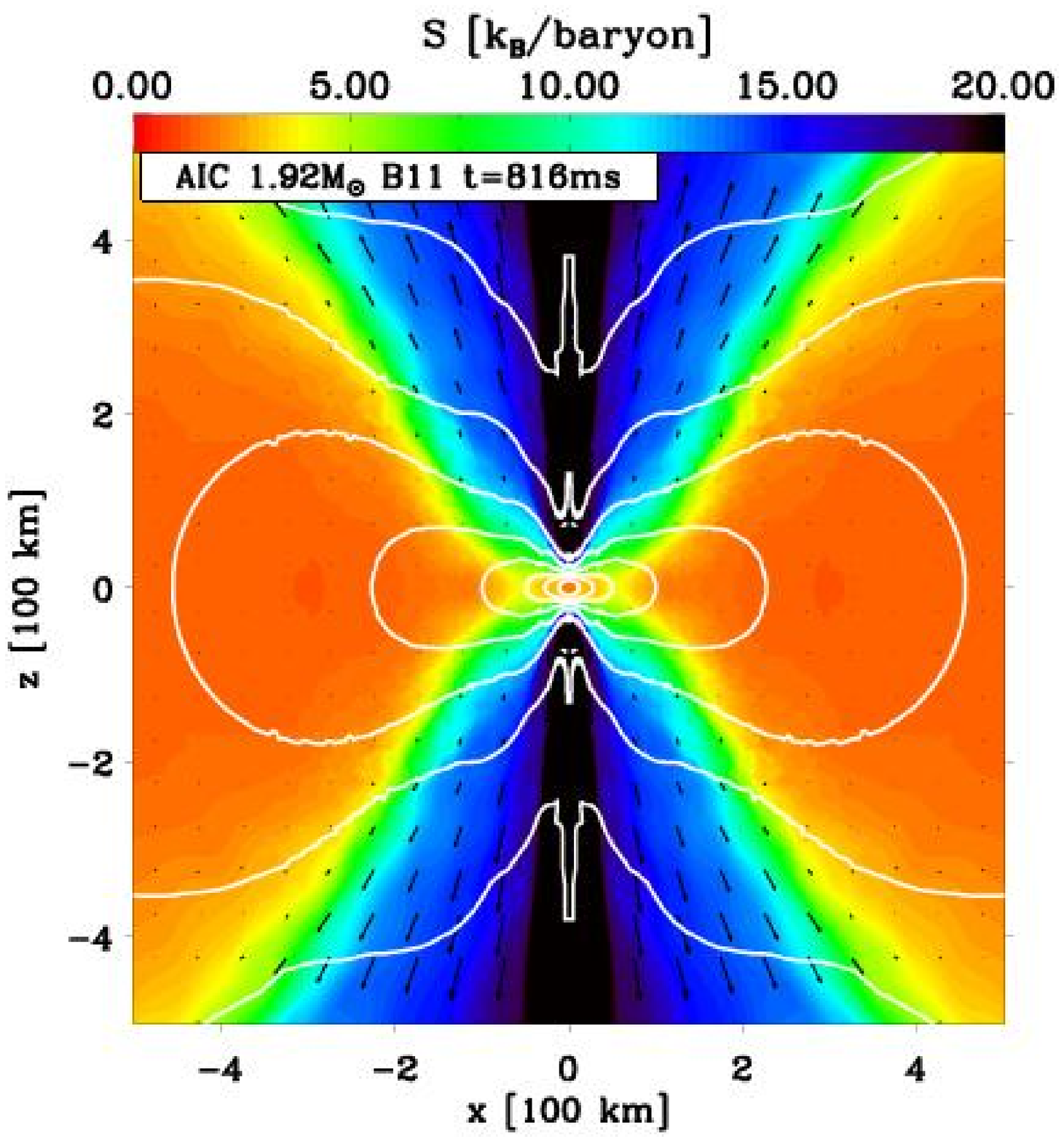}{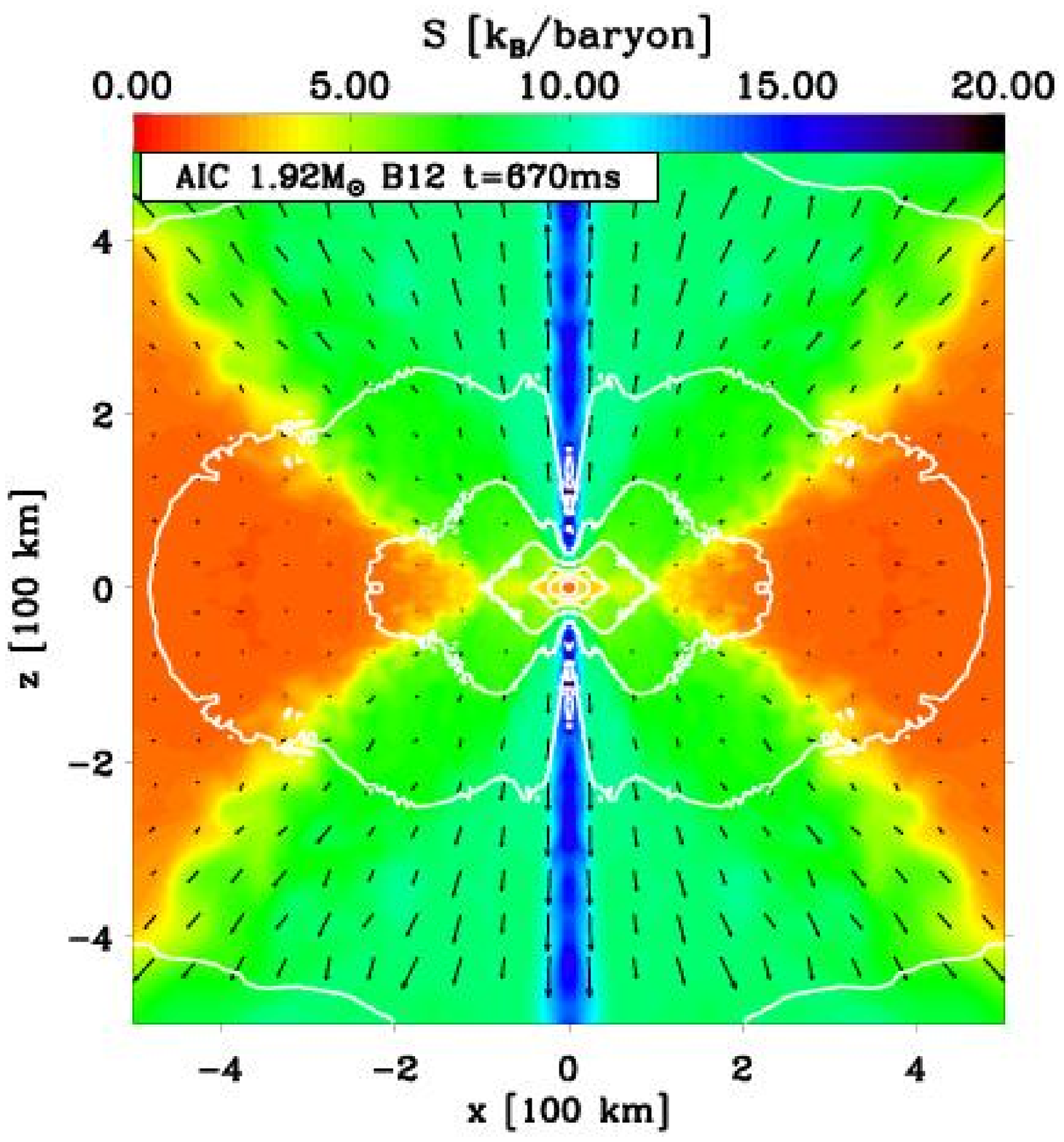}
\plottwo{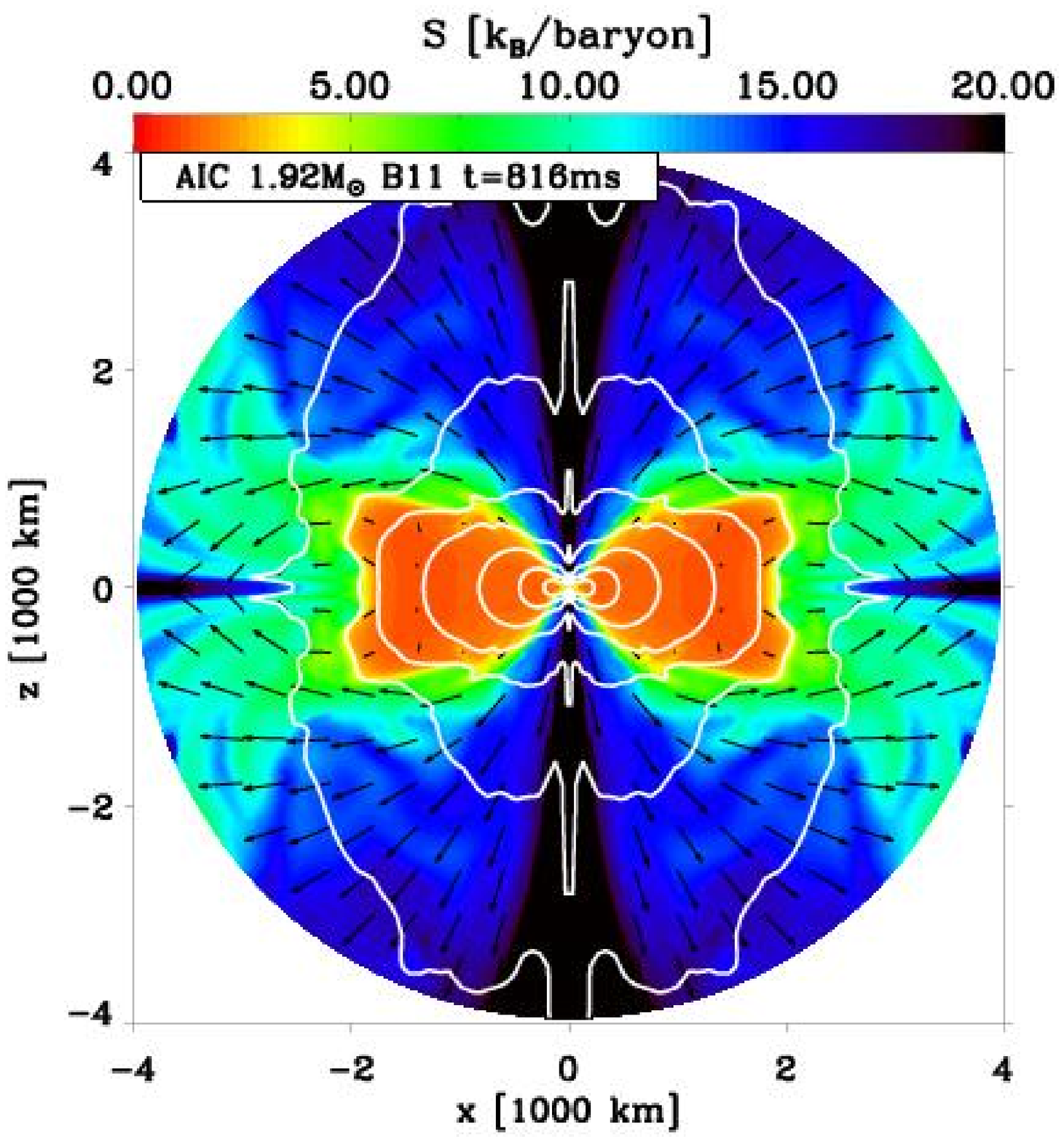}{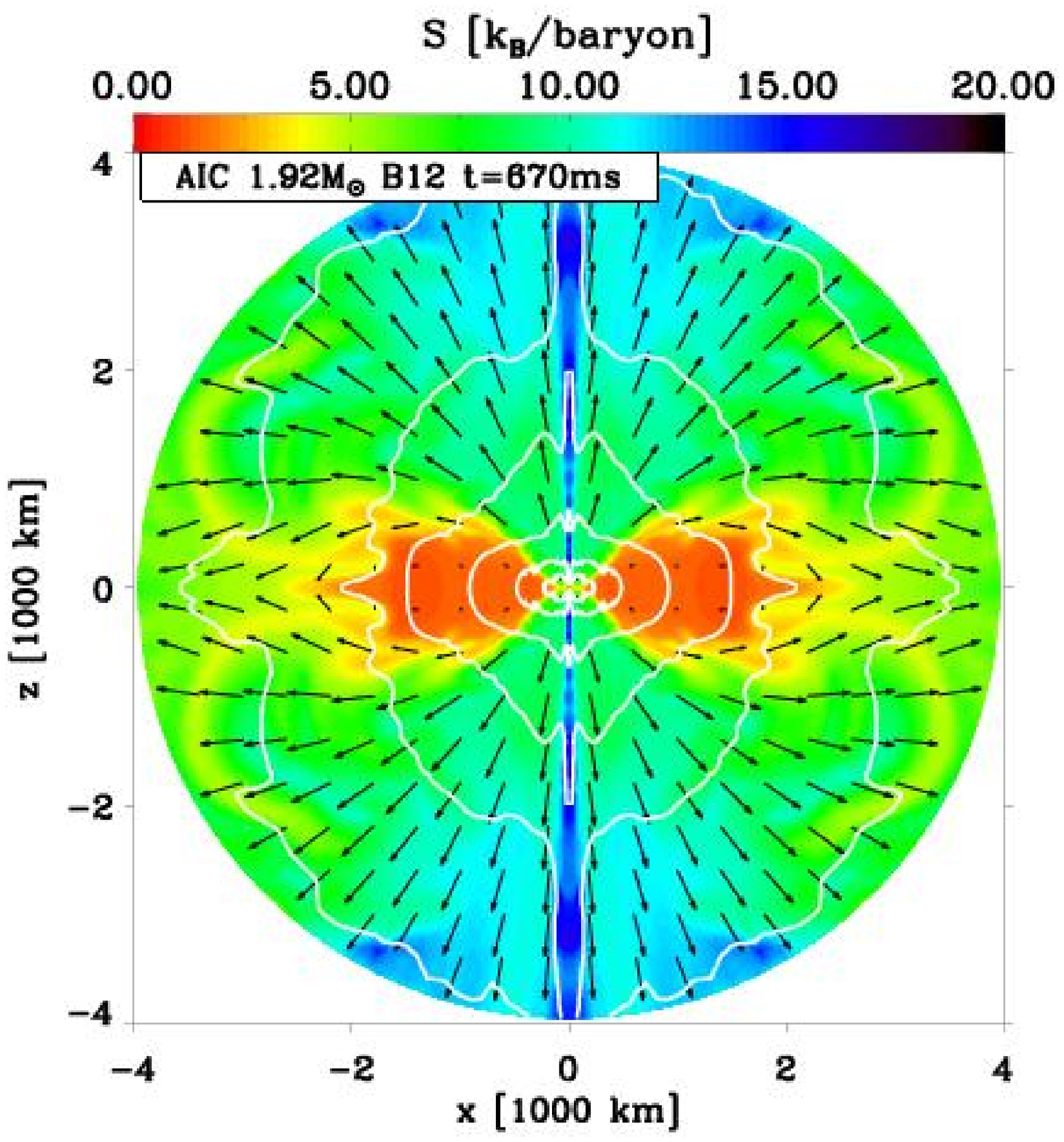}
\caption{
{\it Top:} Colormap of the entropy for the B11 (left) and B12 (right) models, at 816\,ms and
670\,ms after bounce, respectively, and in the inner 1000$\times$1000\,km$^2$, 
using black vectors to represent the velocity field 
(the length is saturated at 5000\,\kms, corresponding to 5\% of the width of the display). 
We also overplot isodensity contours, shown for every decade starting at 10$^{14}$\,g\,cm$^{-3}$.
{\it Bottom:} Same as for the top panels, but extending the maximum displayed radius to
4000\,km, the maximum grid radius in our simulations. The saturation length for the velocity
vectors is now 10000\,\kms. (See text for discussion.)
}
\label{fig_entropy}
\end{figure*}

\section{Description of the explosion}
\label{sect_sim}

   In this section, we present the collapse, the bounce, and the explosion phases of the
1.92\,\mo AIC white dwarf model computed by, and with the method of, Yoon \& Langer (2005). 
In contrast to a similar investigation
performed and analyzed in great detail in D06, magnetic fields are now included, leading to 
little qualitative, but considerable quantitative, changes in the results. 
For completeness, we briefly review the important phases of the explosion
arising from such AIC of white dwarfs. Important results are discussed again, and 
in a broader context, in \S\ref{sect_discussion}.

   Due to the high central density of such white dwarfs 
(above $\sim$10$^{10}$\,g\,cm$^{-3}$, and here set to 5 $\times$ 10$^{10}$\,g\,cm$^{-3}$), 
electron captures on nuclei lead to the collapse of the ONeMg core on a near dynamical timescale.
About 37\,ms after the start of the simulation, the inner core reaches nuclear densities, 
i.e., $\sgreat$2 $\times$ 10$^{14}$\,g\,cm$^{-3}$. A shock wave is born that propagates outward,
but weakens due to the huge energy losses associated with the burst of electron neutrinos and 
the photodissociation of nuclei in the infalling outer-core material. Along the equator, where the
accreting material enjoys sizable centrifugal support, the shock first propagates fast
due to the reduced accretion rate, but eventually stalls due to its sustained magnitude. 
By contrast, along the polar direction where the density profile is much steeper, 
after at first stalling due to the higher accretion rate, the shock soon encounters the layers 
that were formerly at the surface of the white dwarf. As no further accretion occurs, the shock blasts out
of the white dwarf along the pole only 80--100\,ms after bounce and an explosion is initiated. 
Up to that point, the simulations B11 and B12 show the same qualitative and quantitative properties, 
suggesting that magnetic effects on the dynamics are weak or negligible during the first $\sim$100\,ms 
post-bounce evolutionary phases of the AIC. 

\begin{figure*}
\plottwo{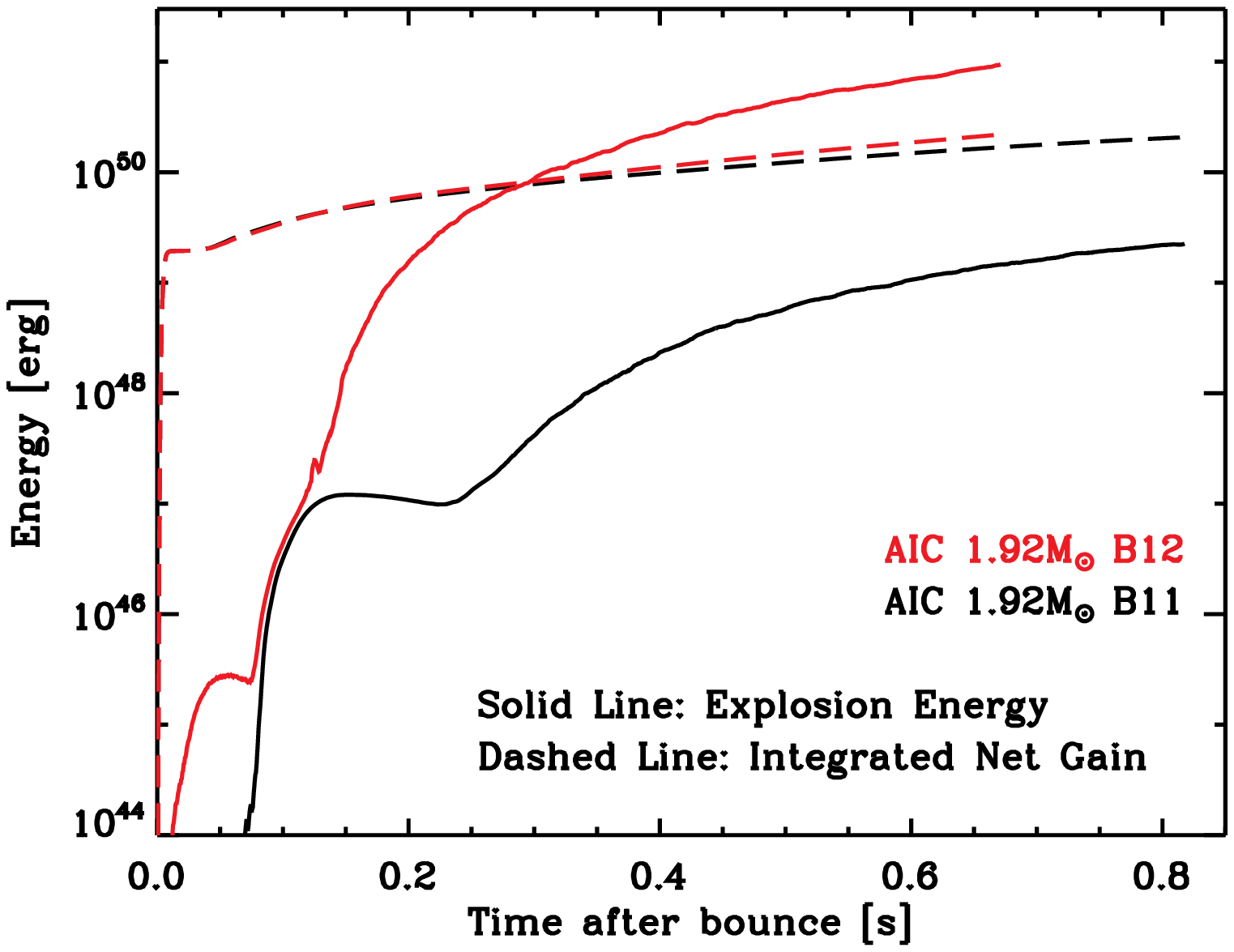}{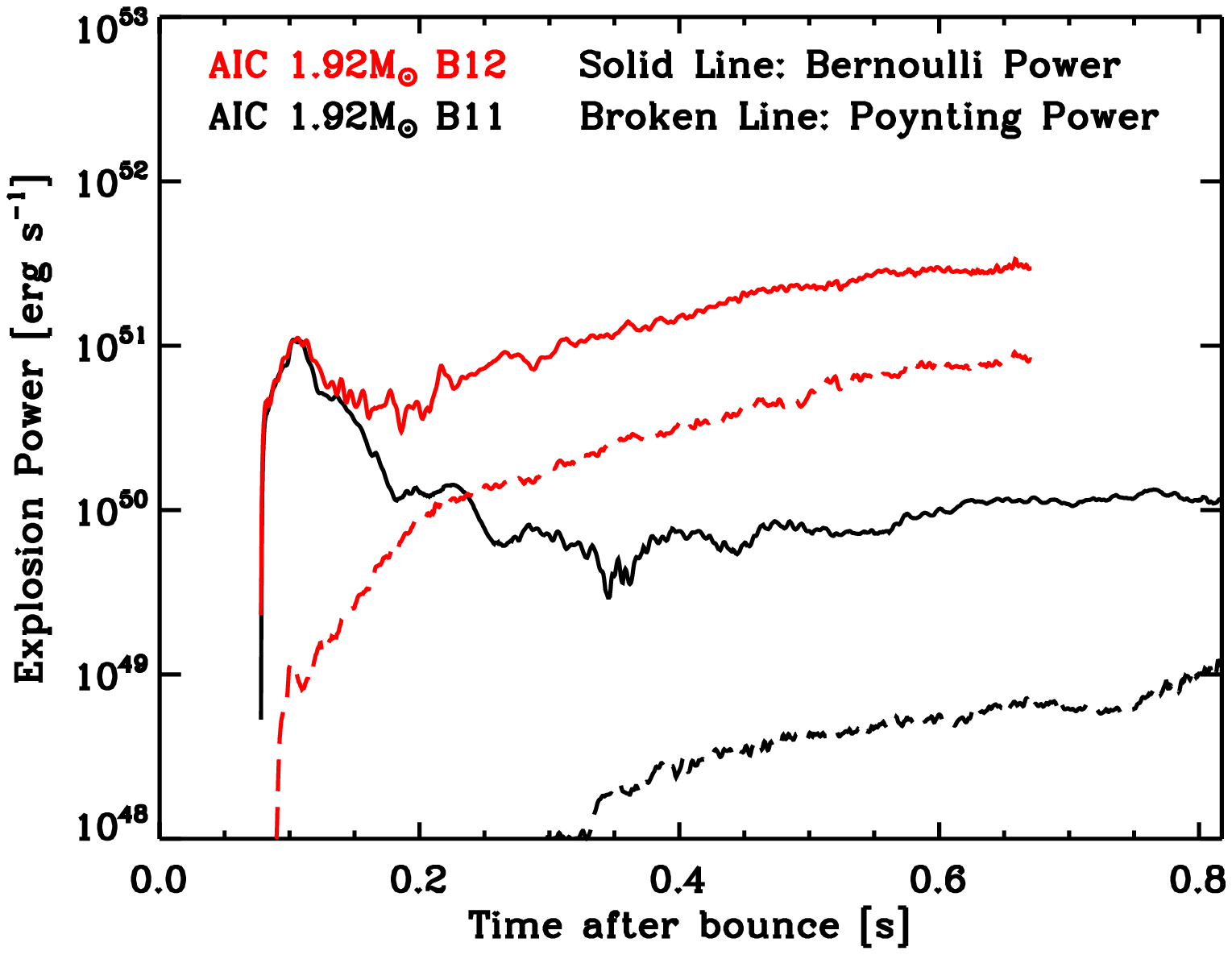}
\caption{
{\it Left:} Time evolution for the 1.92\,\mo model of Yoon \& Langer (2005), with either low (black)
or high (red) initial magnetic fields, of the total energy associated with the ejecta (solid line;
accounting for gravitational, radial and rotational kinetic, thermal, and magnetic energy components)
and the total integrated net neutrino energy gain (broken line). 
{\it Right:} Same as left, but for the Bernoulli
(solid line) and Poynting (broken line) powers in the ejecta, 
computed by integrating the corresponding flux over a radial shell at 500\,km.
}
\label{fig_power}
\end{figure*}

   However, it takes $\sim$100\,ms in the B12 model to amplify the magnetic field
to a value such that magnetic pressure and gas pressure are comparable, at first in the polar
regions at a few tens of kilometers from the protoneutron star (PNS) center.
Compression from the progenitor to a high-density collapsed configuration is the early cause of 
magnetic field amplification, but later it arises from the winding 
of the poloidal field and the continued accretion of the magnetized envelope, spinning up as it 
collapses (see discussion in Burrows et al. 2007b). 
The toroidal field eventually dominates over the poloidal field component
in the inner few hundred kilometers by, typically, 1--2 orders of magnitude.
We speculate that the magnetic properties at $\sim$100\,ms after bounce are comparable
to what would obtain with a fully resolved MRI. 
Note that in our simulations, no significant convection develops behind the shock, as required
for the $\alpha^2$ and the $\alpha$-$\Omega$ dynamo invoked for magnetic-field amplification by
Duncan \& Thompson (1992) and Thompson \& Duncan (1995). 
In any case, in the B12 model after $\sim$100\,ms, the resulting large magnetic pressure 
and associated gradient at the surface of the neutron star power a strong magnetically-driven 
outflow that expands in the previously excavated polar regions of the progenitor white dwarf. 
There is here an analogy with magnetic tower formation in magnetic, differentially-rotating
disks (Lynden-Bell 1996, 2003; Uzdenski \& MacFadyen 2006), since the general picture is one of
magnetic field amplification in a medium that exerts pressure anisotropically.
Here, the presence of excavated polar regions and a dense centrifugally-supported 
fat disk at lower latitudes offers a natural confinement mechanism for the 
magnetically-driven explosion.
Hoop stresses act poorly to confine the blast, since wind particle trajectories originate 
at the neutron star surface over all latitudes between $\sim$40$^{\circ}$ and 90$^{\circ}$\footnote{The analogy 
to an infinitely long thin disk threaded by a magnetic field is not strictly applicable.}.
By comparison, in the B11 model, there is a ``quiescent'' stage after the original blast,  
with some fallback of a fraction of the ejecta. The explosion is revived after $\sim$300\,ms by the
birth of a neutrino-driven wind, as discussed in detail in D06. This wind,
and the neutrino energy deposition that drives it, are the source of the energy of the explosion
in the context of low or negligible magnetic fields. At high B-fields, the neutrinos play only 
a secondary, sub-dominant, role, and then only after a few hundred milliseconds after bounce. 

   We have carried out the B11 and B12 simulations up to 816\,ms and 670\,ms, respectively, 
after bounce.
This constitutes only the very early evolution of the cooling PNS, which typically
takes tens of seconds in total when in isolation (Burrows \& Lattimer 1986). 
We show the entropy (colormaps) and density (contour lines) distributions 
in Fig.~\ref{fig_entropy} for the B11 (left column) and B12 (right column) models 
at the end of each simulation, encompassing the full grid (bottom row) or
the inner region (top row). We can identify three different regions: an inner region where an oblate
neutron star resides, a prolate mass distribution in the ejecta that is mostly confined to polar 
or high-latitude regions (above $\sim$40$^{\circ}$), and a fat and Keplerian disk that closely resembles the
original lobes of the progenitor white dwarf. We now review in detail each component,
together with the global energetics of each explosion.

\begin{figure*}
\plottwo{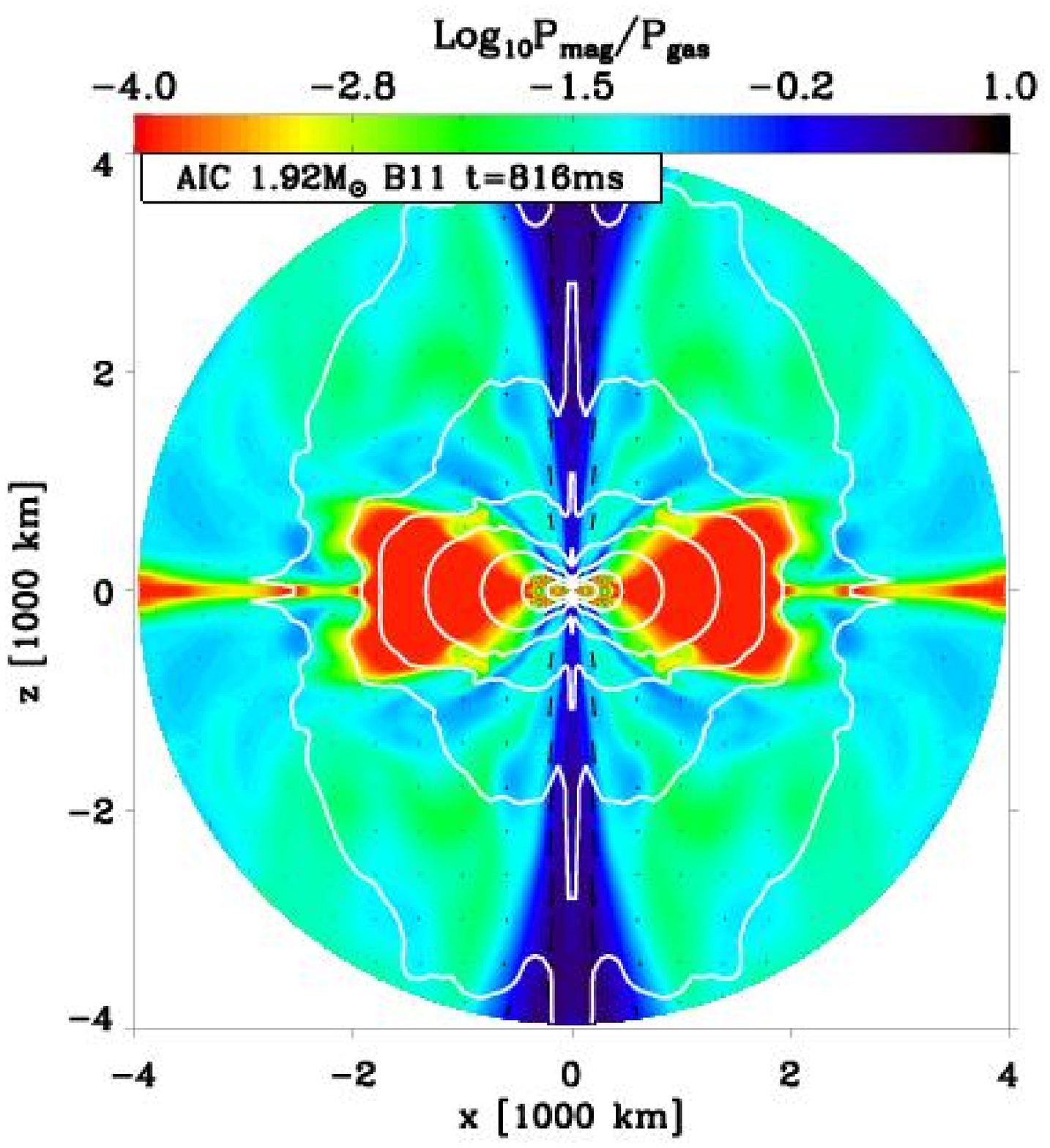}{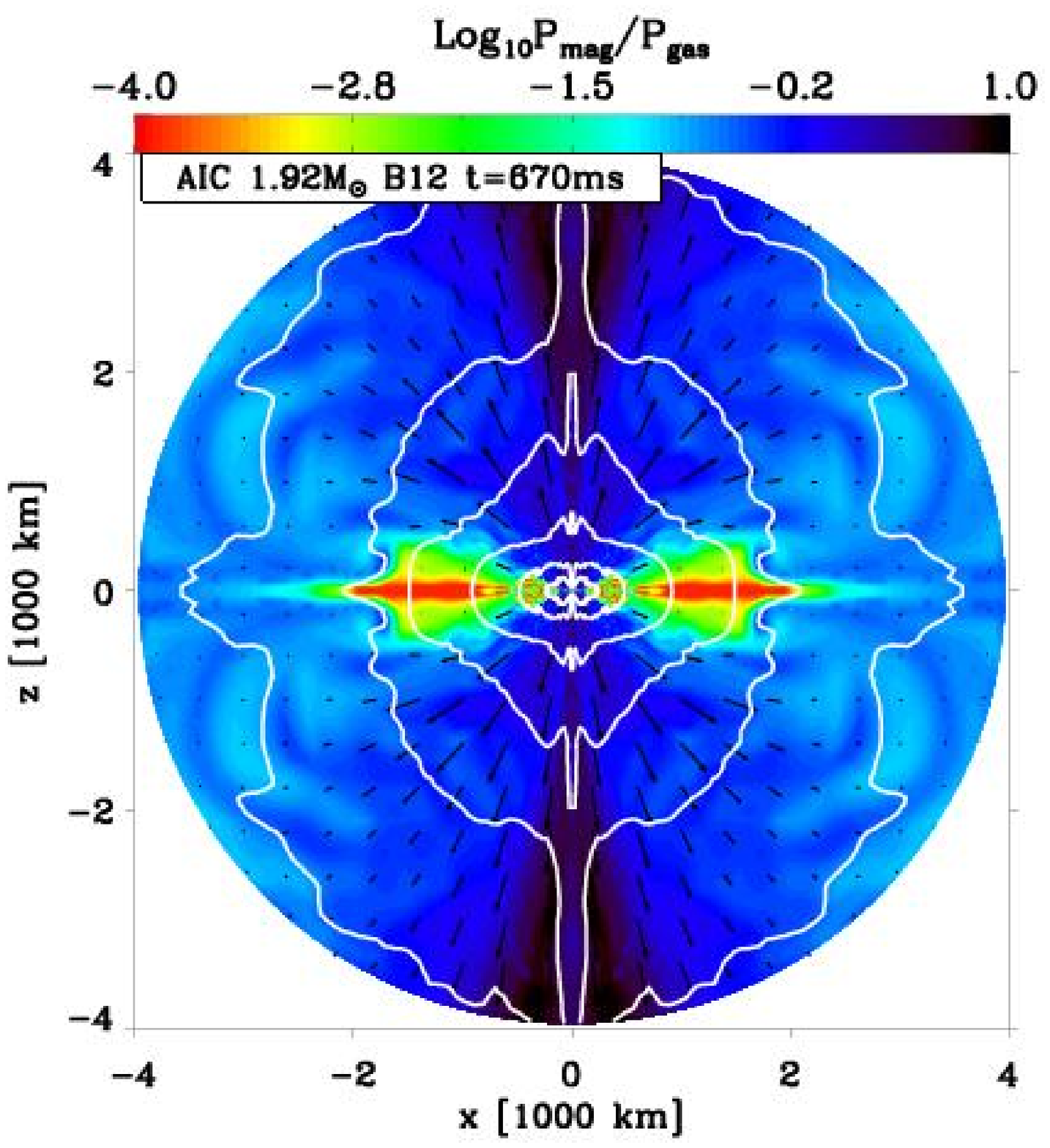}
\plottwo{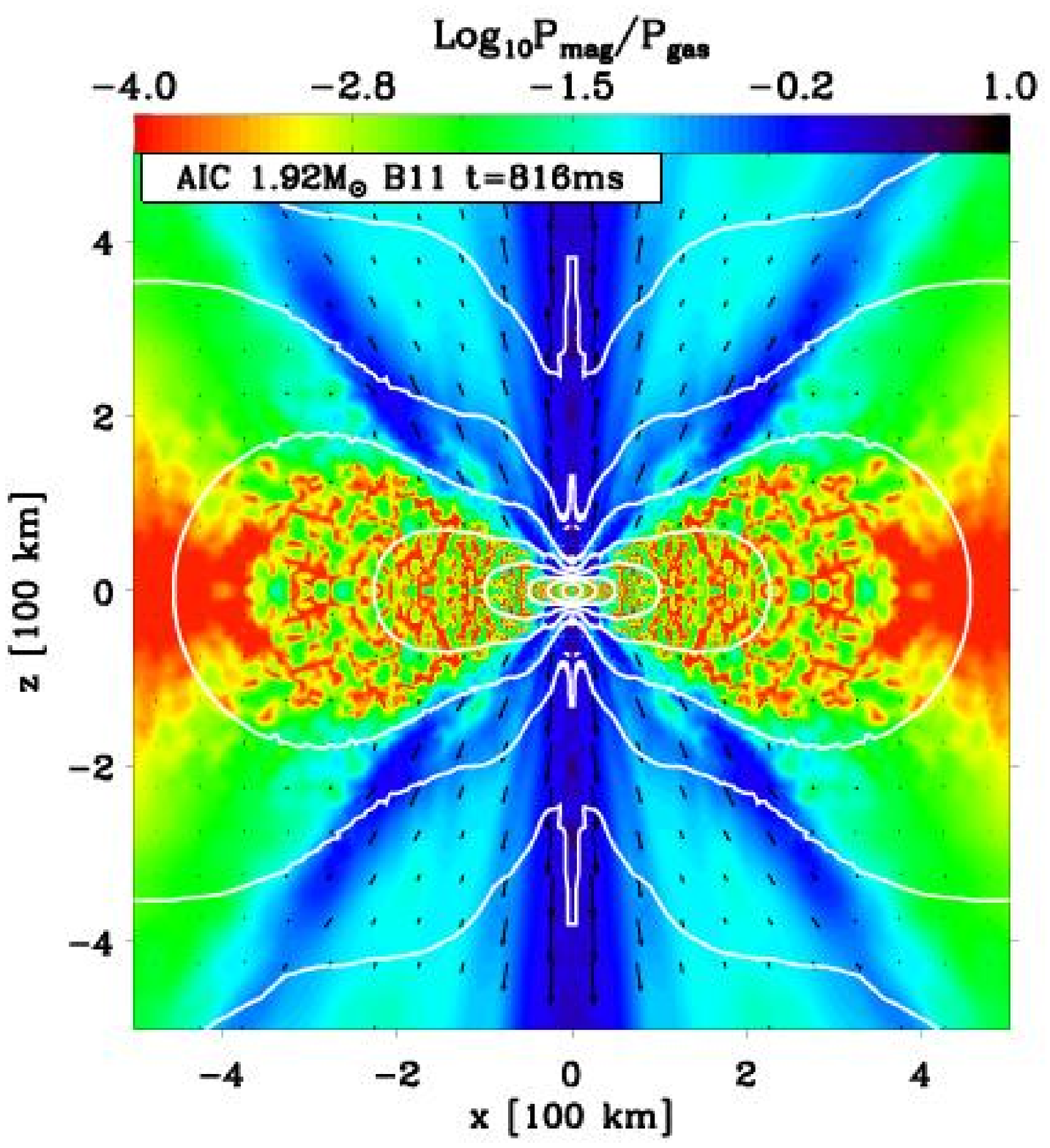}{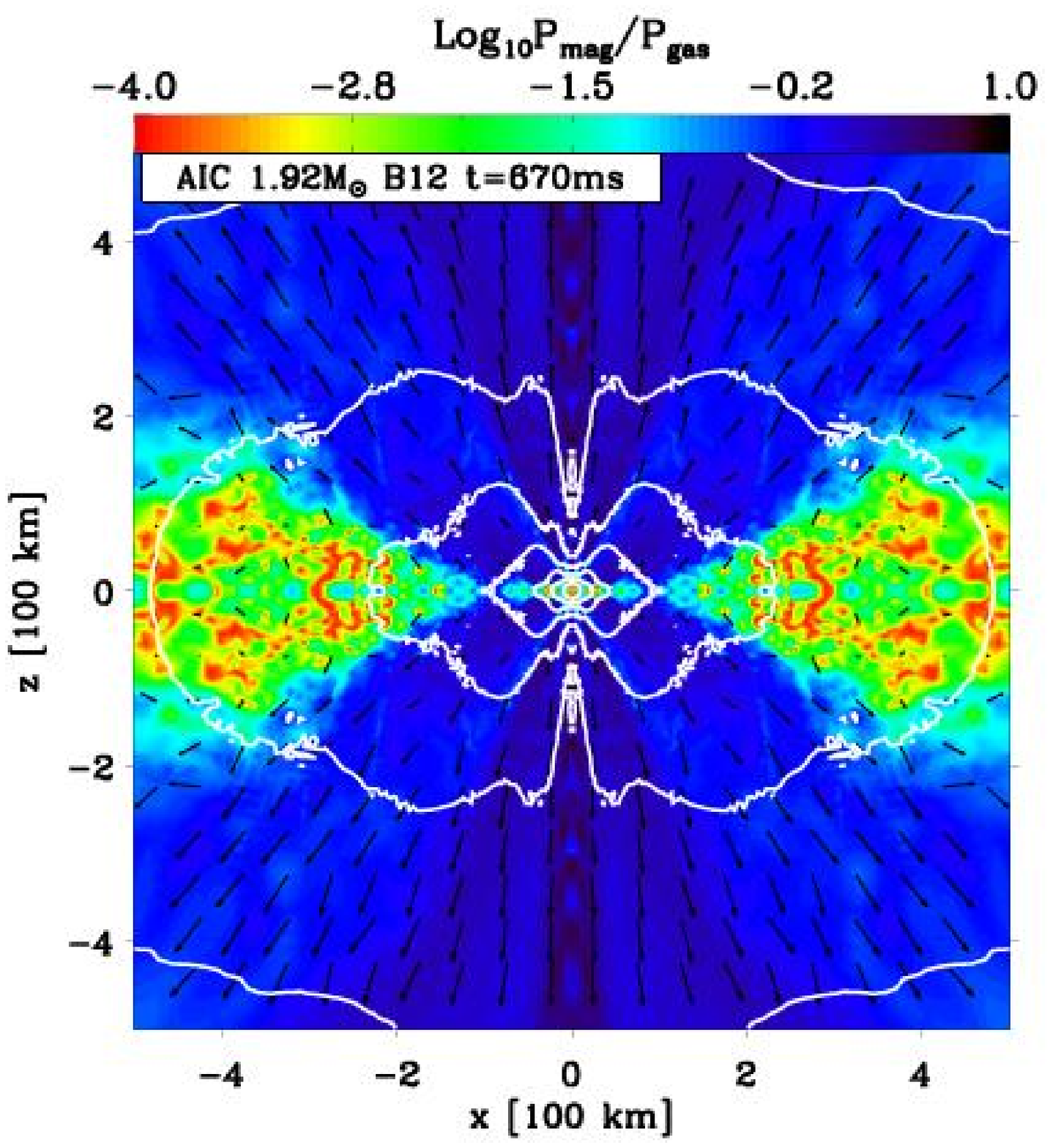}
\caption{{\it Top:} Colormap of the logarithm of the ratio of magnetic pressure to gas pressure 
for the B11 model (left) at 816\,ms after bounce, and for the B12 model (right) at 670\,ms 
after bounce, using black vectors to represent the Poynting luminosity 
(the Poynting flux is multiplied by a factor 4$\pi$R$^2$ and the resulting length is saturated at 
0.1\,B\,s$^{-1}$, which corresponds to 5\% of the width of the display). 
We also overplot isodensity contours, for every decade starting at 10$^{10}$\,g\,cm$^{-3}$, 
which nicely show their change in morphology from oblate in the neutron star vicinity to prolate 
at a few thousand kilometers.  
{\it Bottom:} Same as top, but now zooming in on the inner few hundred kilometers.
The isodensity contours are now plotted up to 10$^{14}$\,g\,cm$^{-3}$.
(See text for discussion.)}
\label{fig_pmpg}
\end{figure*}

\begin{figure*}
\plottwo{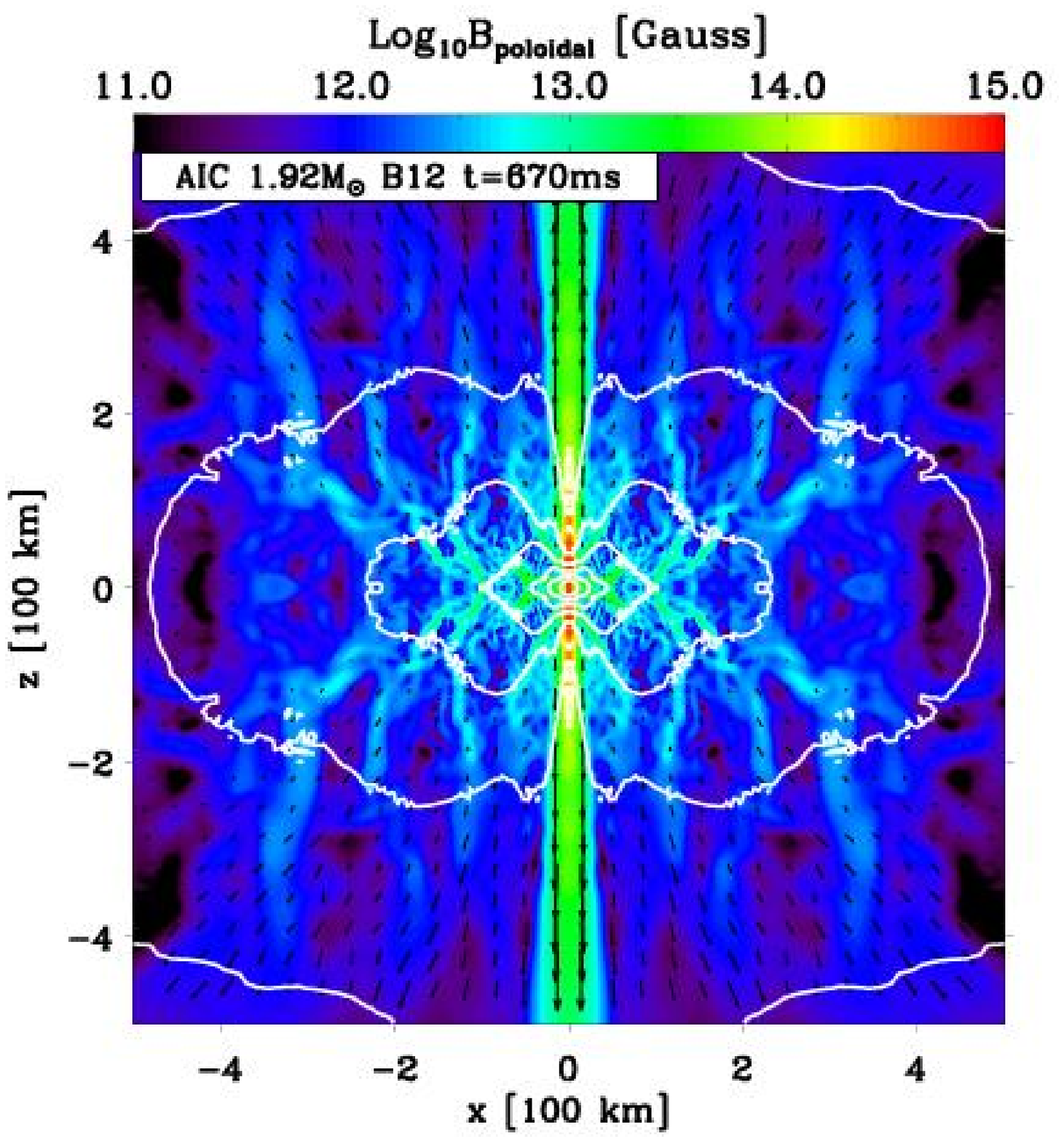}{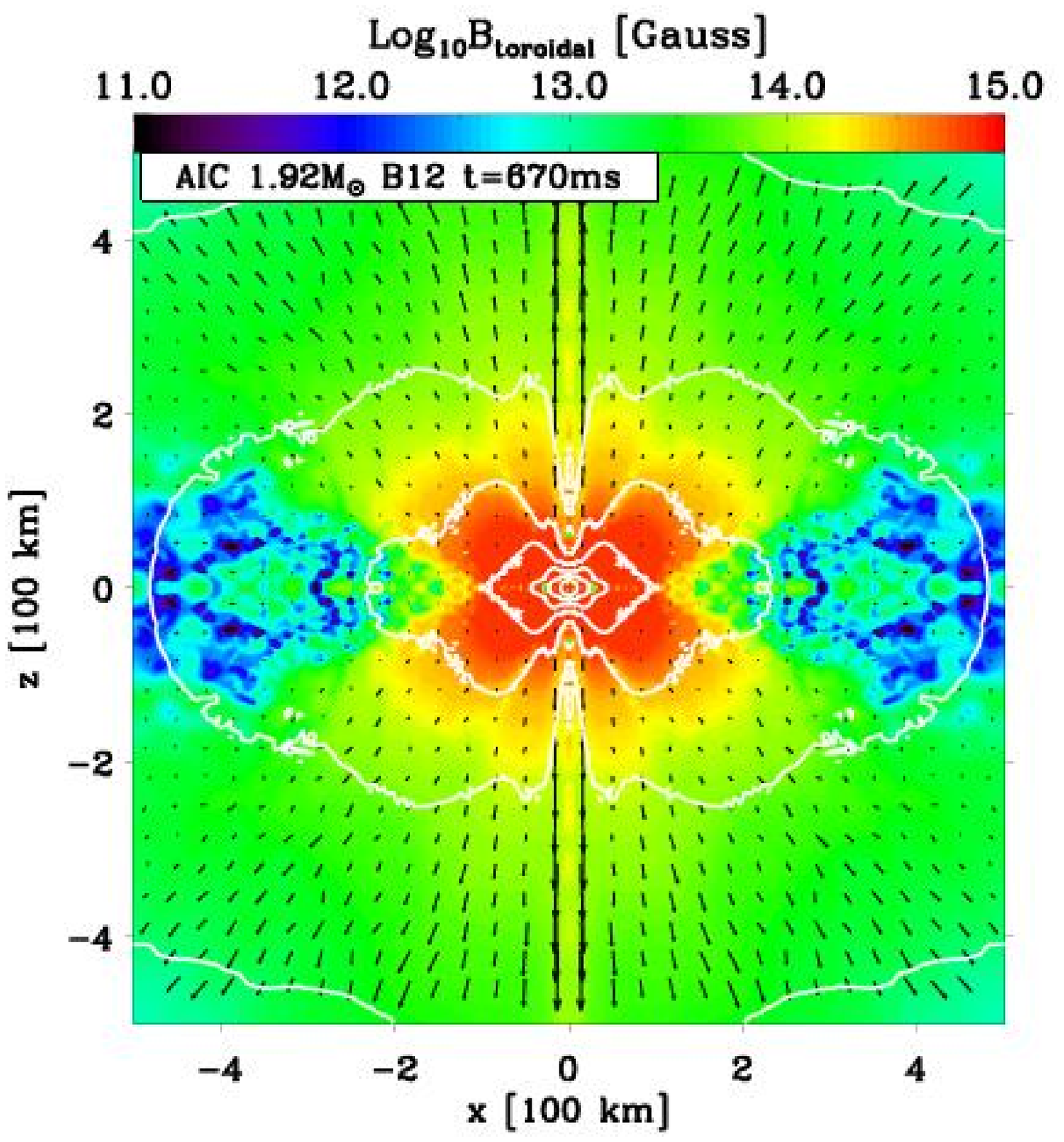}
\caption{{\it Left:} Colormap of the logarithm of the poloidal field magnitude
for the B12 model at 670\,ms after bounce, using black vectors to represent the
velocity (their length is saturated to 7\% of the width of the diplay for a
magnitude of 10000\,\kms).
We also overplot isodensity contours, for every decade starting at 10$^{14}$\,g\,cm$^{-3}$
{\it Right:} Same as left, but for the logarithm of the toroidal field magnitude.
(See text for discussion.)}
\label{fig_bpbt}
\end{figure*}

In the left panel of Fig.~\ref{fig_power}, we show the time evolution, for both models (B12: 
Red; B11: Black), of the explosion energy (solid line) and integrated net neutrino
energy gain as a function of time after bounce (we limit this integration to regions 
with a mass density less than 10$^{10}$\,g\,cm$^{-3}$; broken line). 
With high initial B-field, the
explosion energy reaches $\sim$1\,B, two orders of magnitude higher than in the
corresponding low B-field case (which explodes under-energetically, but easily, nonetheless), 
and one order of magnitude
larger than the integrated net gain through neutrino energy deposition. At low B-field,
the neutrinos play a critical role in powering the explosion, but the explosion is
symptomatically weak, on the order of 0.01\,B. 
In the right panel of Fig.\ref{fig_power}, we show the 
power injected into the ejecta, associated both with the 
Bernoulli\footnote{The Bernoulli flux is
defined by the quantity $\rho V_R \times (e_{\rm th} + e_{\rm kin} + e_{\rm rot} + P/\rho - G M_{\rm PNS} / R)$, 
where $V_R$ is the radial velocity, $e_{\rm th}$ is the thermal energy, $e_{\rm kin}$ and $e_{\rm rot}$ 
are the ($r,z$) and rotational kinetic energies, $G$ the gravitational constant, $M_{\rm PNS}$ the PNS mass 
(taken as the cumulative mass of material with a density greater than 10$^{10}$\,g\,cm$^{-3}$), and 
$R$ the spherical radius. Both the Bernoulli flux and the Poynting flux are then integrated through a sphere
at 500\,km to give a power, or a luminosity (Fig.~\ref{fig_power}).} 
flux (solid line) and the 
Poynting flux (broken line) integrated through a sphere at 500\,km. Although the Poynting flux decreases
roughly as the inverse square of the radius above the neutron star surface, it is larger 
in the B12 simulation than the Bernoulli flux in the B11 simulation.
Combined with the strong Bernoulli flux, the explosion power reaches $\sim$2\,B\,s$^{-1}$.
 
   This gain in explosion energy stems partly from the extraction of the free energy of rotation 
of the PNS, the difference between the kinetic energy of rotation and the corresponding
energy for solid-body rotation at the same total PNS angular momentum.
Selecting the regions inside of a density cut of 10$^{8}$, 10$^{10}$, 
or 10$^{12}$\,g\,cm$^{-3}$, the
total rotational energy difference between the B11 and the B12 cases is on the order
of $\sim$10\,B. While a large fraction of this energy is used to expand the PNS and, therefore,
do work against gravity (where it is the most costly, i.e., at the base of the potential well),
some of this energy is made available to the explosion and the growth of the magnetic field,
whose associated gradient at the PNS surface is responsible for driving mass outward at 
the escape speed (Shibata et al. 2006; Burrows et al. 2007b). 
In the B12 simulation, the PNS density structure is significantly modified,
with isodensity contours more spatially separated, and with less contrast between the 
polar and equatorial radii (see the top row of Fig.~\ref{fig_entropy} for an illustration).
Magnetic support is partly responsible for the more extended PNS configuration.
As a result, the PNS in the B12 model suffers significant spin-down, another indication that
rotational energy has been extracted from the compact object by magnetic torques. 
The average period in the inner 10$^{10}$\,g\,cm$^{-3}$ increases from 15\,ms to 21\,ms,
and in the inner 10$^{12}$\,g\,cm$^{-3}$ increases from 2\,ms to 2.5\,ms, for models
B11 and B12, respectively.

\begin{figure*}
\plottwo{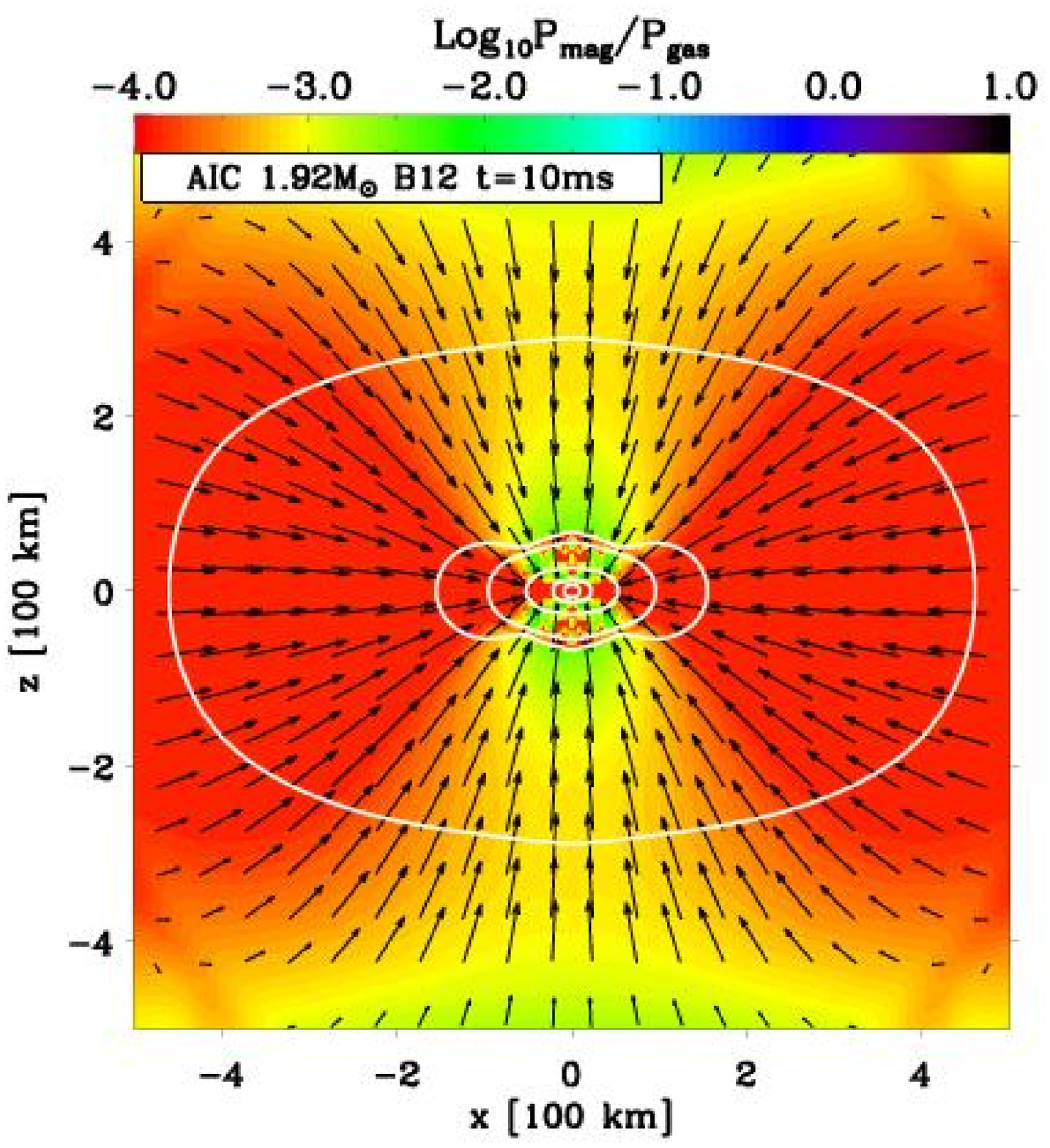}{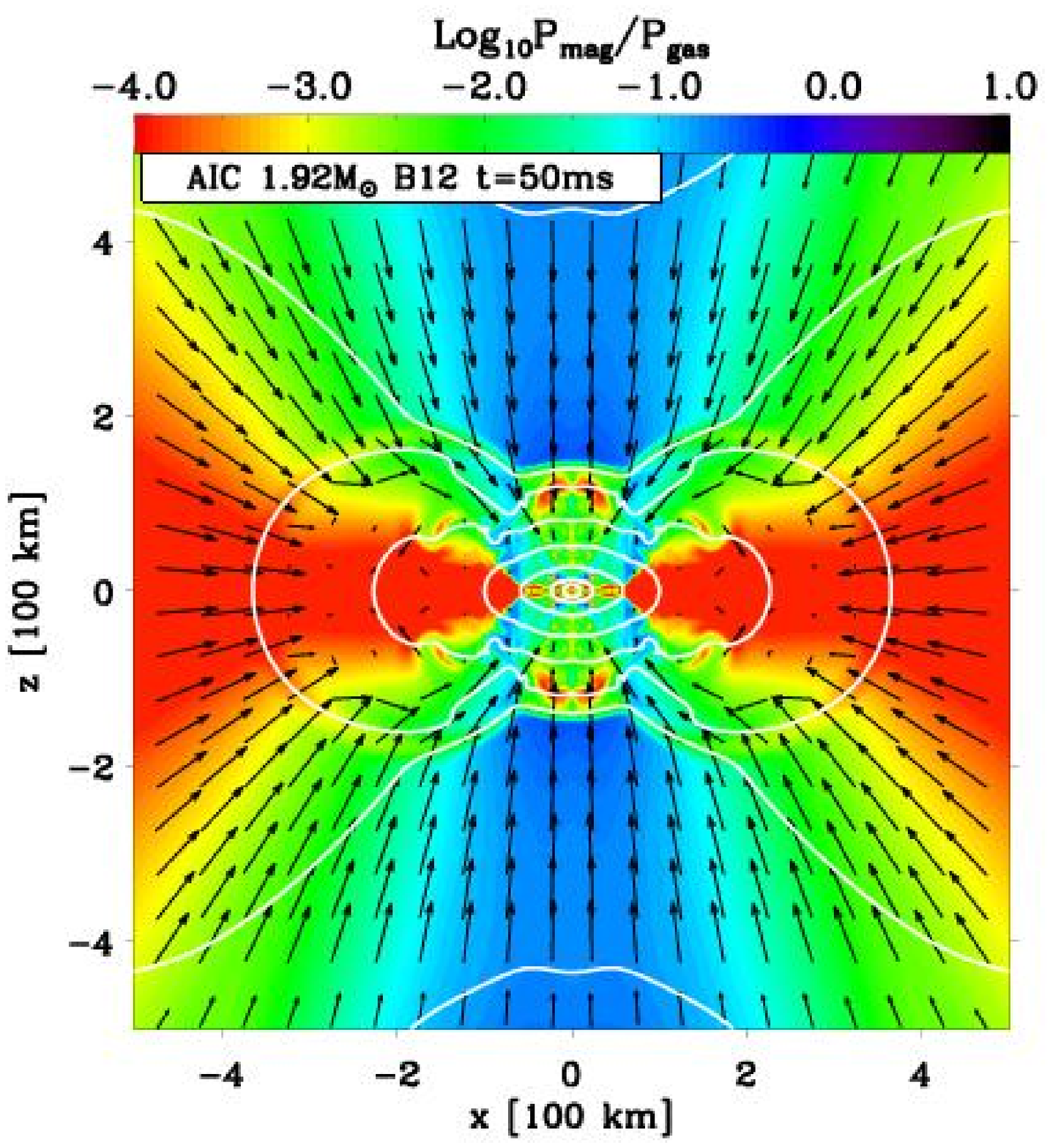}
\plottwo{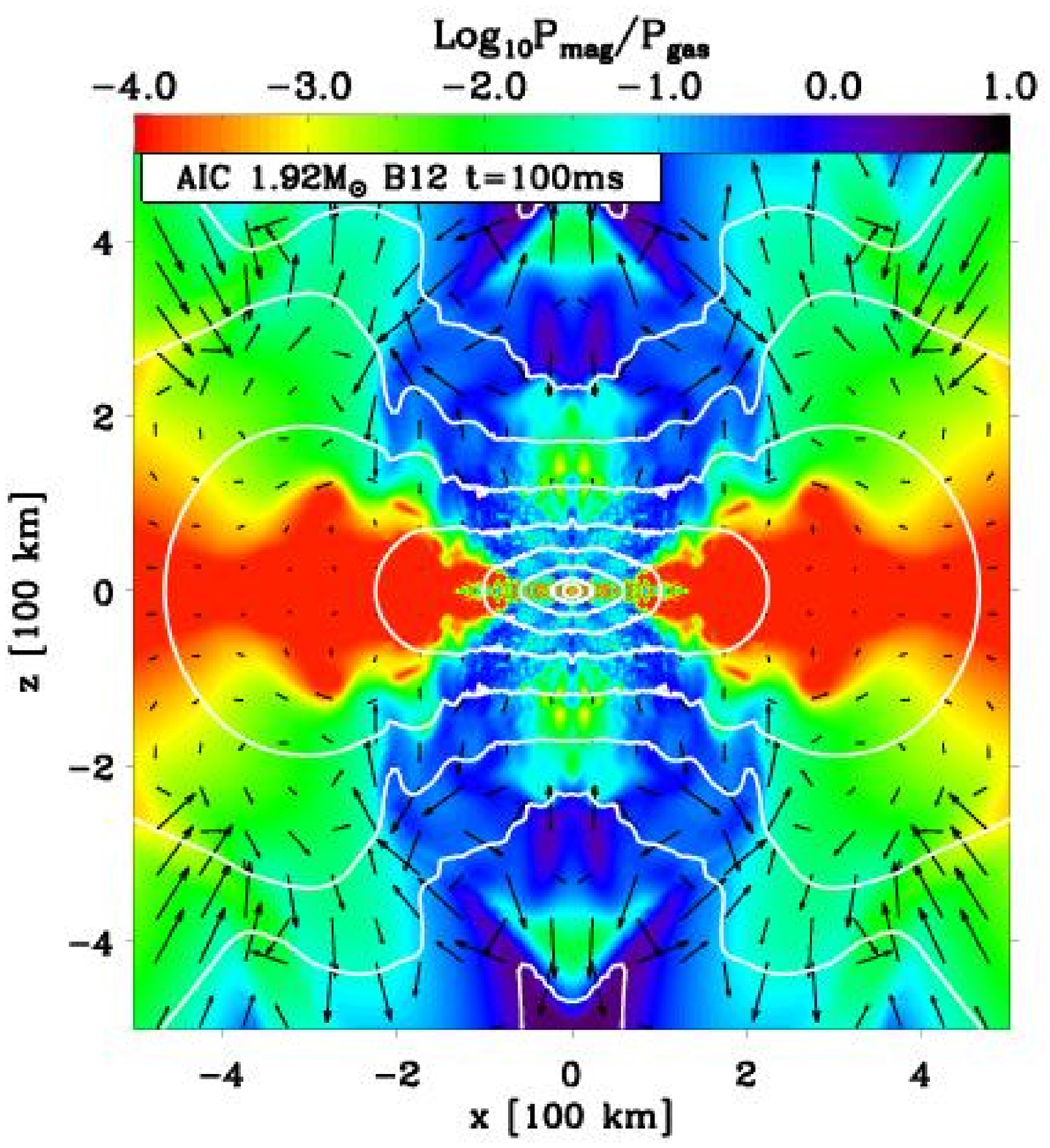}{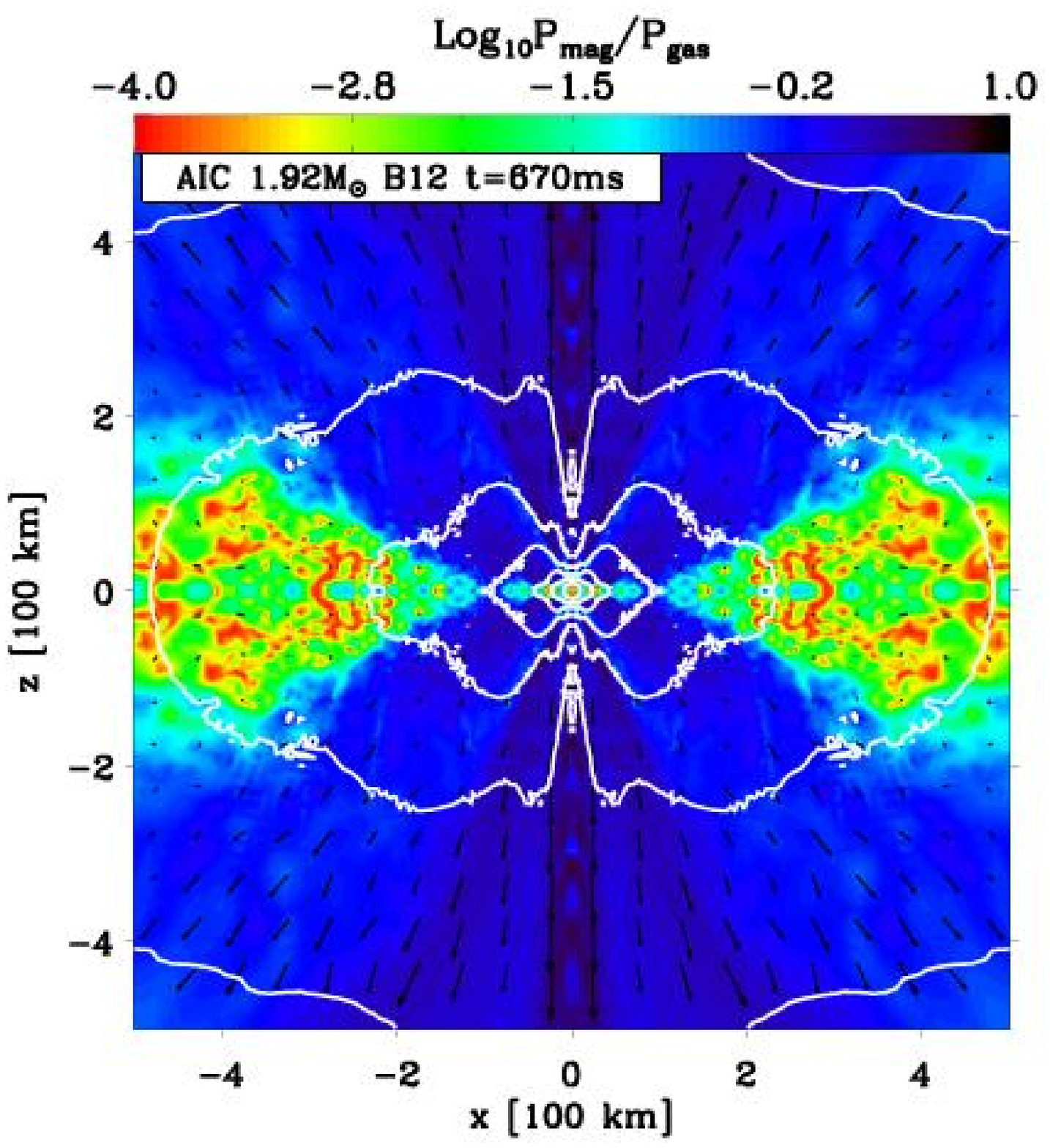}
\caption{Colormap of the logarithm of the ratio of magnetic pressure to gas pressure for the B12 
model, at 10\,ms (top left), 50\,ms (top right), 100\,ms (bottom left), and 670\,ms 
(bottom right) after bounce, using white isodensity contours starting 
at 10$^{14}$\,g\,cm$^{-3}$ shown for every decade in density, as well as 
black vectors to represent the velocity, saturated at a length of 7\% of the width of 
the display for a magnitude of 5000\kms.
}
\label{fig_pmpg_seq}
\end{figure*}

In Fig.~\ref{fig_pmpg}, we illustrate the magnetic properties at the last computed time 
in each model, for both large and small scales, by showing colormaps of the ratio of the
magnetic pressure to the gas pressure.
These are over-plotted with vectors depicting the Poynting luminosity (flux scaled
by the quantity $4 \pi R^2$, where $R$ is the spherical radius).
Note how in the B12 model the magnetic pressure is large and comparable to the gas pressure 
in most of the ejecta. 
The ejecta retain a higher Poynting flux along the polar direction,
even at large distances of a few thousand kilometers, and this is likely caused by the significant 
confinement of the inner jet. At mid-latitudes, the Poynting flux drops faster than the inverse
square of the radius.
The only excluded regions, where the magnetic field
has no substantial effect, are the neutron star core (layers with a local density greater than
10$^{12}$\,g\,cm$^{-3}$)  and the lobes of the progenitor white dwarf (low-density, non-collapsed
regions of the progenitor white dwarf).
In these latter regions, characterized by large vorticity but little bulk radial advection, 
we notice large fluctuations in the magnetic to gas pressure ratio, 
stemming mostly from fluctuations in the magnetic field energy. 
Although we cannot identify strong convection, stabilized by the differential rotation,
moderate velocity shear can induce such local enhancements of the magnetic field.
In Fig.~\ref{fig_bpbt}, we present the spatial distribution in the inner 500\,km of each
of the poloidal (left) and toroidal (right) components of the magnetic field for the B12 model
at 670\,ms after bounce.
Unable to resolve the MRI, we predict the dominance of the toroidal component of the magnetic
field over the poloidal component, typically by about two orders of magnitude, with the exception
of the axis region where it is the strong radial advection of material that
generates a strong poloidal field
from the original strong toroidal field of the corresponding mass parcel (in the B11 model,
equipartition is reached in this region, but the gas density and pressure are two orders of
magnitude lower than in the same region in the B12 model). While the relative
poloidal and toroidal field magnitudes produced by an operating MRI would likely differ,
the total magnetic energy we obtain may agree approximately, as it is
fundamentally extracted from the (given) free energy of rotation available in the
differentially rotating surface layers of the PNS.
As in Fig.~\ref{fig_pmpg}, we show in Fig.~\ref{fig_pmpg_seq} the ratio of the
magnetic pressure to the gas pressure, but at 10\,ms, 50\,ms, 100\,ms,
and 670\,ms after bounce. This figure portrays how the pressure is more and more magnetically-dominated 
directly outside the neutron star; the magnetically-driven wind advects highly-magnetized 
material to large distances.

\begin{figure}
\plotone{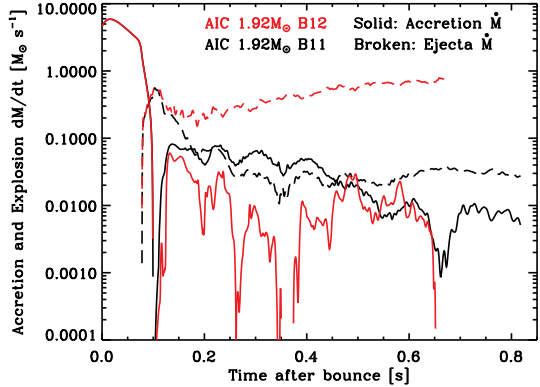}
\caption{Time evolution for the B11 (black) and B12 (red) models of the instantaneous integrated  
mass flux accreting (solid) or outflowing (broken) through a shell at a radius of 500\,km.
(See text for discussion.)}
\label{fig_mdot}
\end{figure}

\begin{figure*}
\plottwo{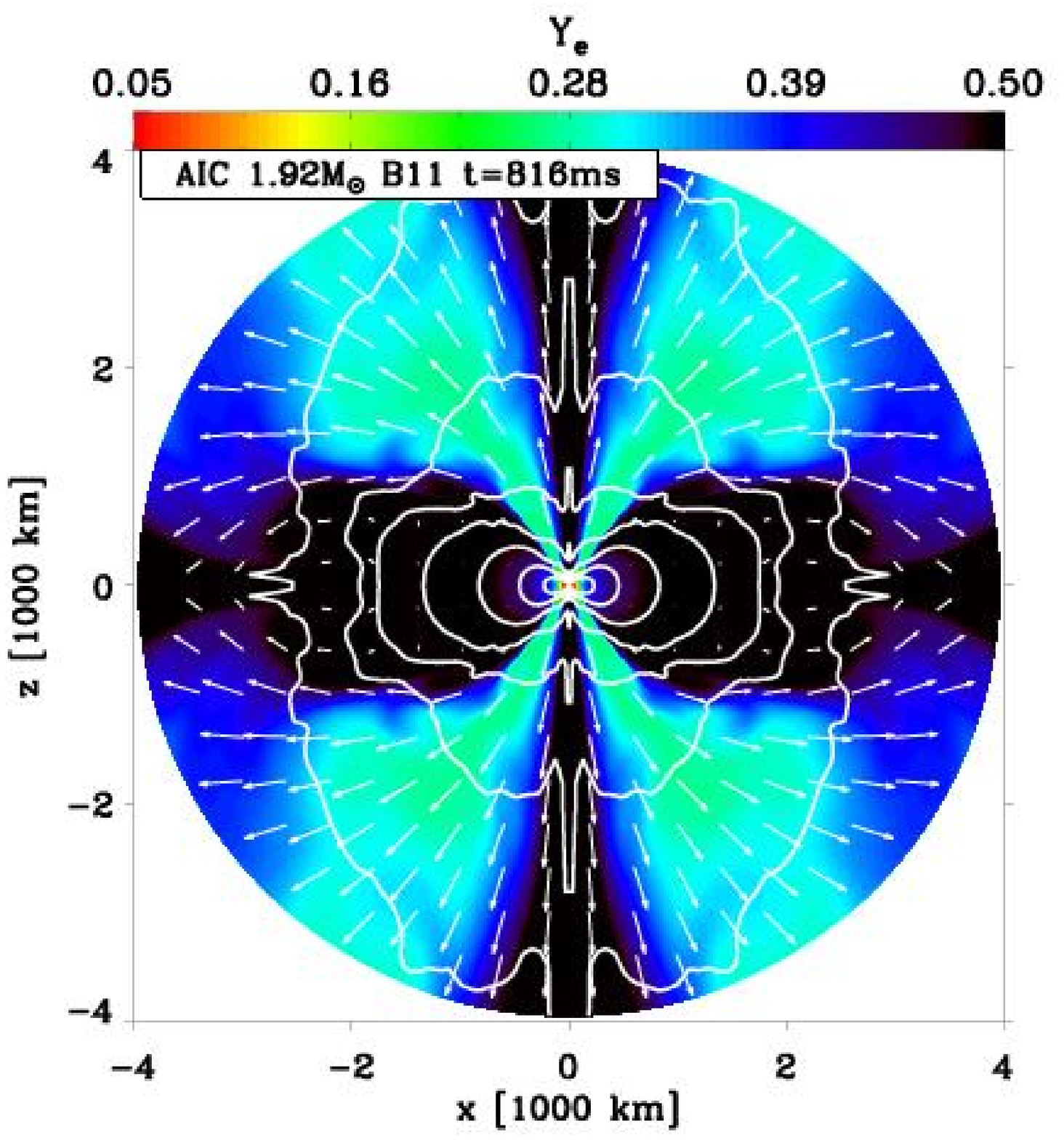}{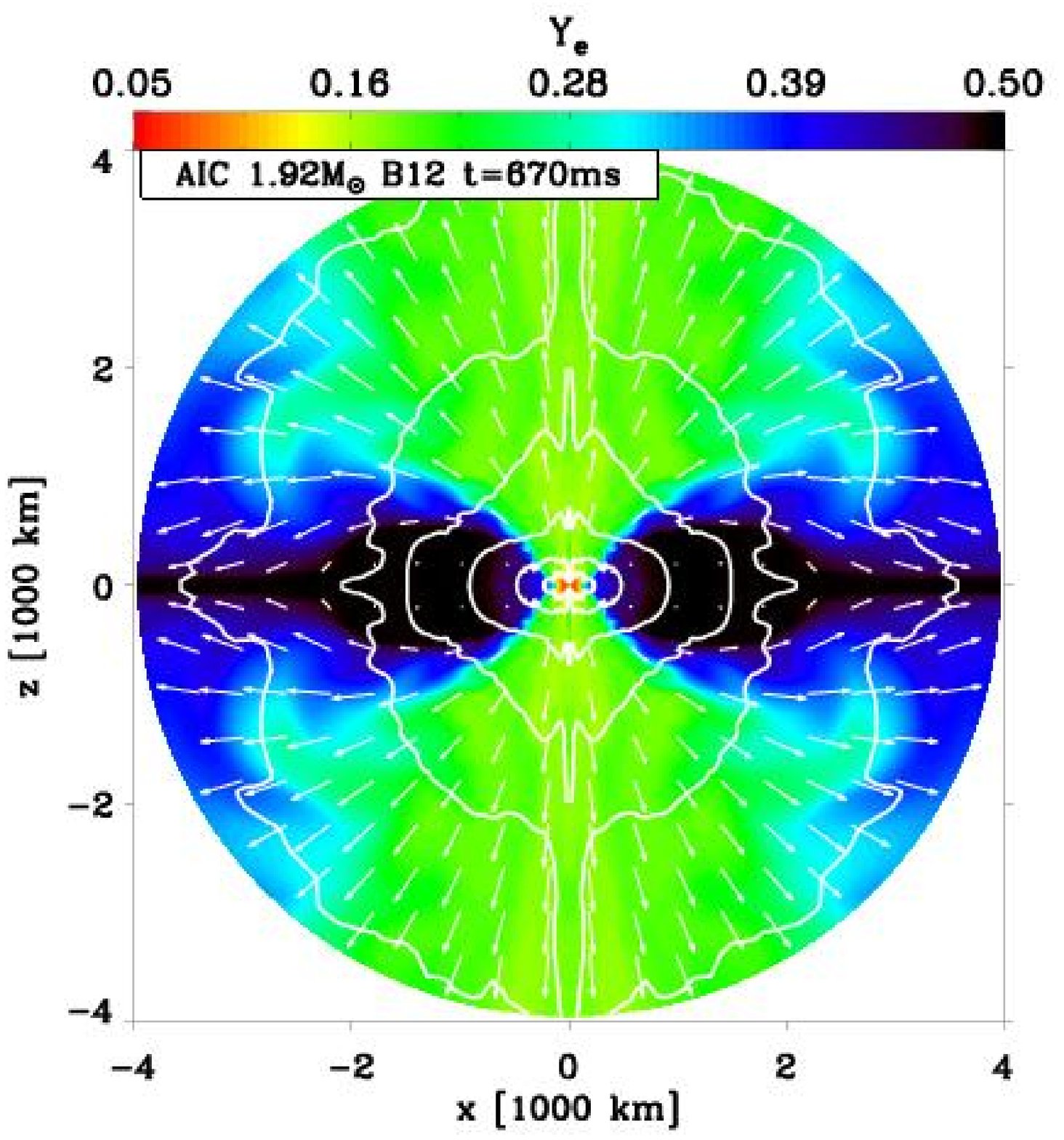}
\plottwo{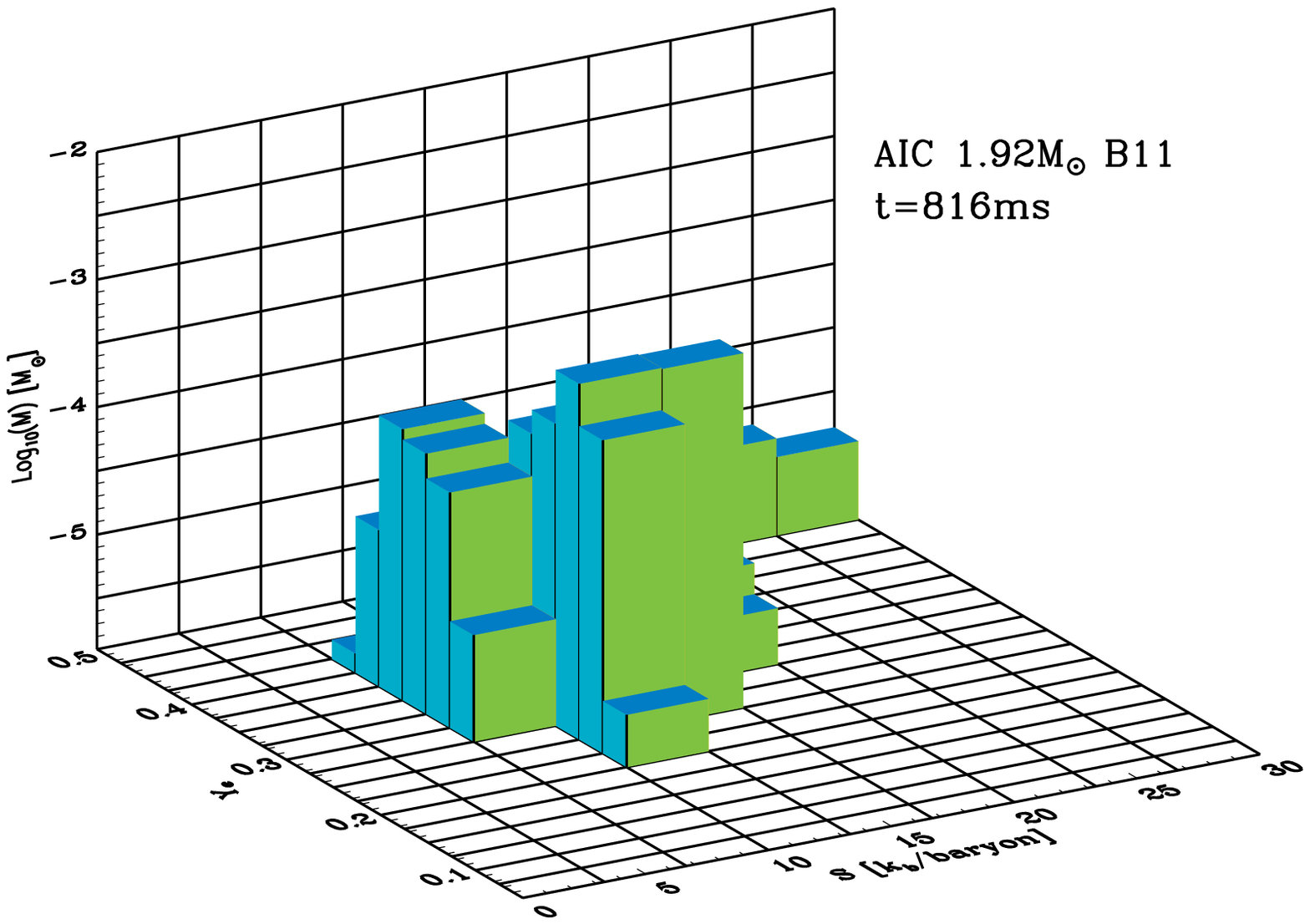}{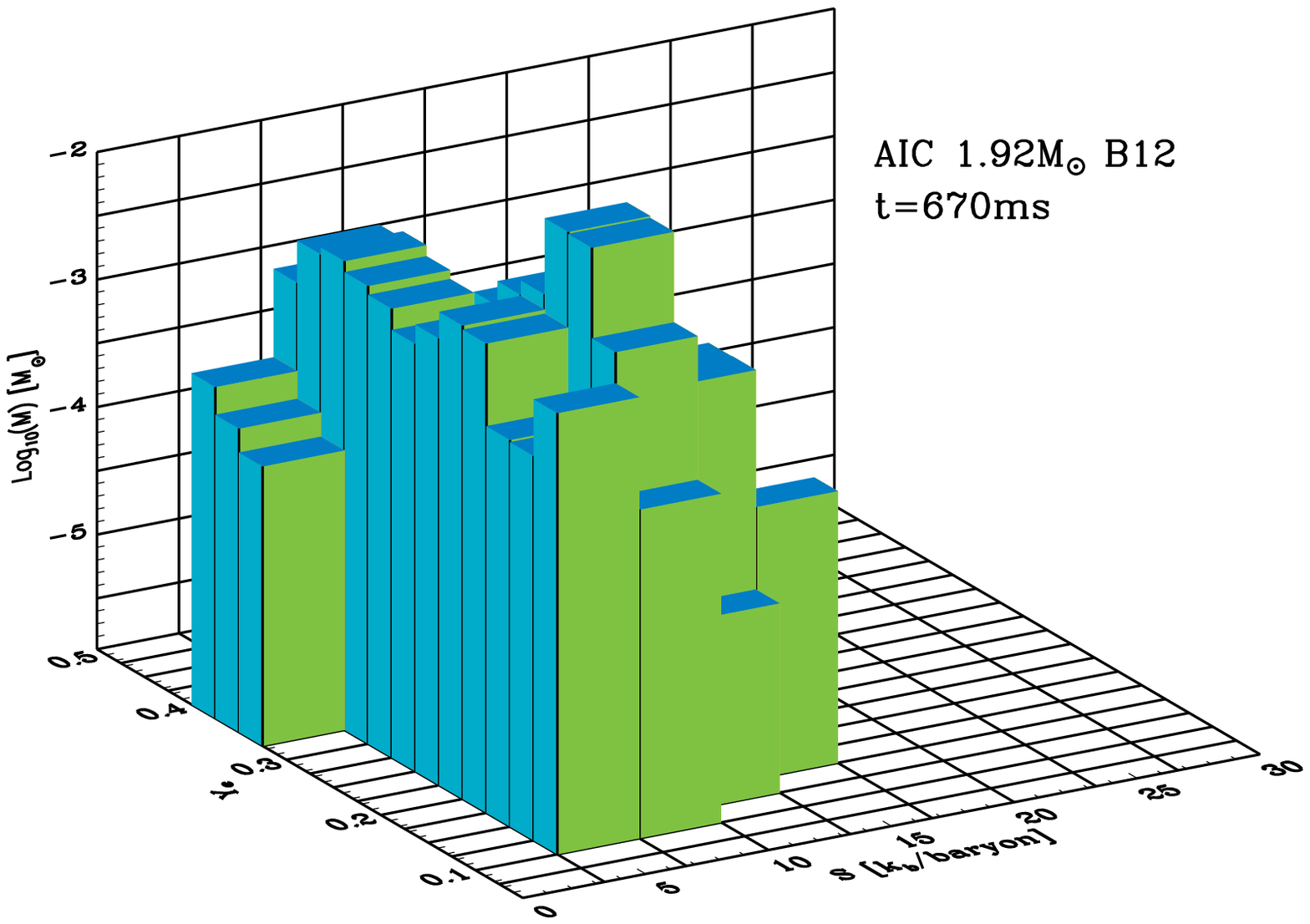}
\caption{
{\it Top:} Colormap of the electron fraction $Y_{\rm e}$ at the end of the simulation
for model B11 (left) and model B12 (right), with white isodensity contours for the density 
shown every decade starting at 10$^{10}$\,g\,cm$^{-3}$, and white velocity contours
saturated at 7\% of the width of the display for a magnitude of 10000\kms. 
{\it Bottom:} Three-dimensional histogram of the distribution of the ejecta mass as a function of 
electron fraction, $Y_{\rm e}$, and entropy, $S$, for the B11 (left) and B12 (right) models.
Notice the larger total ejecta mass, and the stretching of the electron-fraction distribution to 
values as low as 0.1, in the B12 model. (See text for discussion.)}
\label{fig_ye_dist}
\end{figure*}

   Mass loss in the explosion is considerably enhanced in the B12 model, the 
cumulative mass expelled being two orders of magnitude larger than in the B11 model,
i.e., 0.1\,\mo compared with a few $\times$ 0.001\,\mo (Fig.~\ref{fig_mdot}). 
In both cases, accretion in the equatorial 
regions, where material is centrifugally supported, is of the order of 0.01\,\mo\,s$^{-1}$.
In the low B-field case, this is nearly compensated by the mass lost through
the neutrino-driven wind. In the B12 model, the mass loss rate through the
magnetically-driven wind is nearly two orders of magnitude larger, at 0.7\,\mo\,s$^{-1}$,
and the nascent neutron star actually loses mass. Interior to a density cut of 
10$^{10}$\,g\,cm$^{-3}$, the PNS baryonic mass is 1.32\,\mo for the B12 model,
compared with 1.47\,\mo for the B11 model (both at $\sim$500\,ms after bounce).
At the end of the B12 simulation and at the maximum radius of 4000\,km, the ejecta density
is 10$^{5}$--10$^{6}$\,g\,cm$^{-3}$ and its velocity is 40000\,\kms.
This is about two orders of magnitude denser and 20\% slower than the neutrino-driven wind we obtain at
comparable times in the B11 model. Hence, while the ejecta are magnetically driven,
they are not Poynting-flux dominated and they are even more non-relativistic.

  In D06, we found that the electron fraction of the ejected material
was bimodal, with symmetric material of high entropy (i.e., $\sim$25\,$k_{\rm B}$\,baryon$^{-1}$) 
ejected along the poles, and neutron-rich material of lower entropy 
(i.e., $\sim$15\,$k_{\rm B}$\,baryon$^{-1}$) at mid-latitudes. Here, the electron fraction is 
clustered around a neutron-rich mean, with $Y_{\rm e} = 0.25$ and an entropy of 
$\sim$8\,$k_{\rm B}$\,baryon$^{-1}$.  
We illustrate these results in Fig.~\ref{fig_ye_dist} with histograms for the electron 
fraction and entropy distributions of the ejecta material.
Given the $\sim$0.1\,\mo of material expelled  in the B12 model with such 
characteristics, these yields may have some significance for the composition of the 
interstellar medium in r-process nuclei (see \S\ref{sect_ye}), as suggested by Wheeler et al. (1998).
 
%
%
%
%
%
%
%
%

  The long-term development of the explosion is influenced by the early excavation
of the polar regions of the white dwarf. Hence, the explosion remains confined to the
poles, extending only down to a $\sim$40$^{\circ}$ latitude. Some fraction of the material
on the polar-facing side of the excavated white dwarf is entrained by the ejecta, the velocity
shear at the interface inducing Kelvin-Helmholtz rolls that mix low and high $Y_{\rm e}$ material
(the lower resolution of our simulations in the region of interest, around a few hundred kilometers,
underresolves this instability, but some rolls are clearly seen).
At the end of the simulation, this interface has visibly been eroded and the lobes of the white dwarf
progenitor subtend a smaller opening angle (compare the left and right panels in the top row
of Fig.~\ref{fig_entropy}). 
Interestingly, most of this entrained material does not reach the escape speed from this AIC progenitor,
of the order of 15000\,\kms, and ultimately remains bound to the system. In practice, its trajectory
curves back towards the equatorial plane, which it hits at a radius of 2000-3000\,km from the PNS.
Hence, material with low $Y_{\rm e}$ is being fed into the backside of the disk, from regions 
that were originally near the pole\footnote{Note that, initially, in our Eulerian approach, we fill the
ambient medium around the progenitor white dwarf with low-density, low-temperature material,
which, nonetheless, may do work against the ejecta expansion and lead to the fall back of some 
material at the outskirts of the disk. At the start of the simulation, this ambient pressure 
$P_{\rm ambient}$ is about 8$\times$10$^{19}$\,dyne\,cm$^{-2}$, over four order of magnitude 
smaller than the pressure in the surface layers of the white dwarf. The $PdV$ work done by 
the ejecta to expel this ambient material translates into an energy penalty on the order of 
$P_{\rm ambient} \times ( \frac{4}{3} \pi R_{\rm max}^3 - \pi R_{\rm eq}^2 \times 2 R_{\rm pol} )
\sim 2 \times 10^{46}$\,erg, where $R_{\rm max}$ is the maximum radius of the grid, 
and $R_{\rm eq}$ and  $R_{\rm pol}$ are the equatorial and polar radii of the progenitor white dwarf.
This energy penalty is small in comparison with the final energy of the explosion.}.
If the explosion persists over a long time, such ablation may completely erode and 
destroy the original lobes of the progenitor white dwarf, although perhaps 
it may merely redistribute this material to larger distances still within the white dwarf 
potential well. Material located beyond a distance of $\sim$1000\,km from the PNS
and above a $\sim$40$^{\circ}$ latitude follows an essentially radial, ballistic, trajectory
and escapes the gravitational potential of the massive white dwarf.

At the end of both simulations, the neutron stars have a strongly oblate structure and rotate
fast (see also the comparable equilibrium configurations of Liu \& Lindblom 2001). 
Although the pole-to-equator radius contrast is essentially
preserved, in the B12 model the 10$^{10}$\,g\,cm$^{-3}$ isodensity contour, 
for example, is stretched out to $\sim$100\,km at mid-latitudes, 
and to only $\sim$50\,km in the B11 model
(compare the contour morphology in the left and right panels, bottom row, of Fig.~\ref{fig_entropy}).
The same holds for the temperature, which remains large along the pole even at 100\,km
($\sim$3\,MeV in the B12 model compared with $\sim$1\,MeV in the B11 model).
Higher magnetic fields modify the hydrostatic configuration of the neutron star, leading 
to its spatial extension, and the subsequent peeling back of the corresponding regions by the 
magnetically-driven wind. 

\begin{figure*}
\plottwo{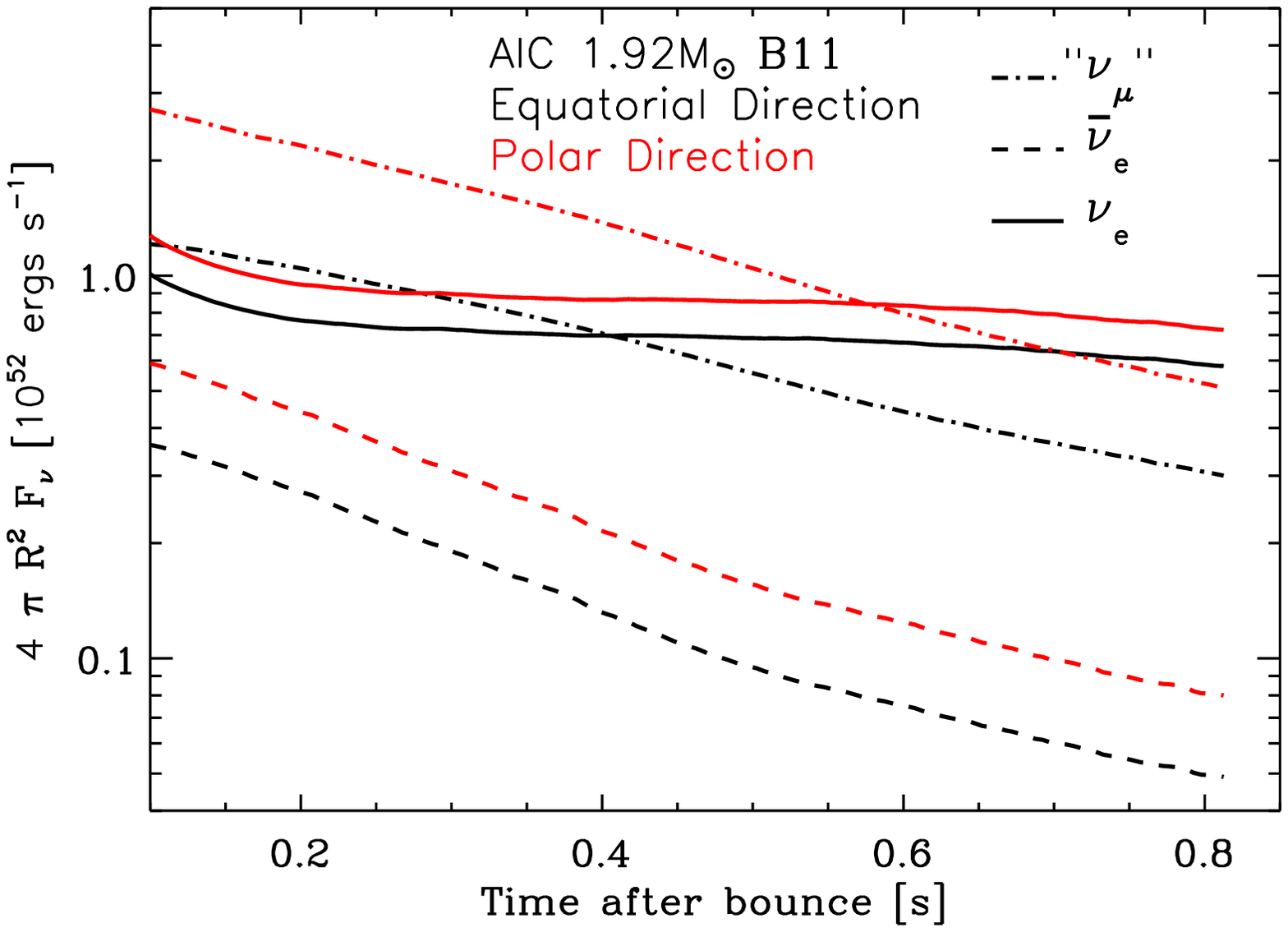}{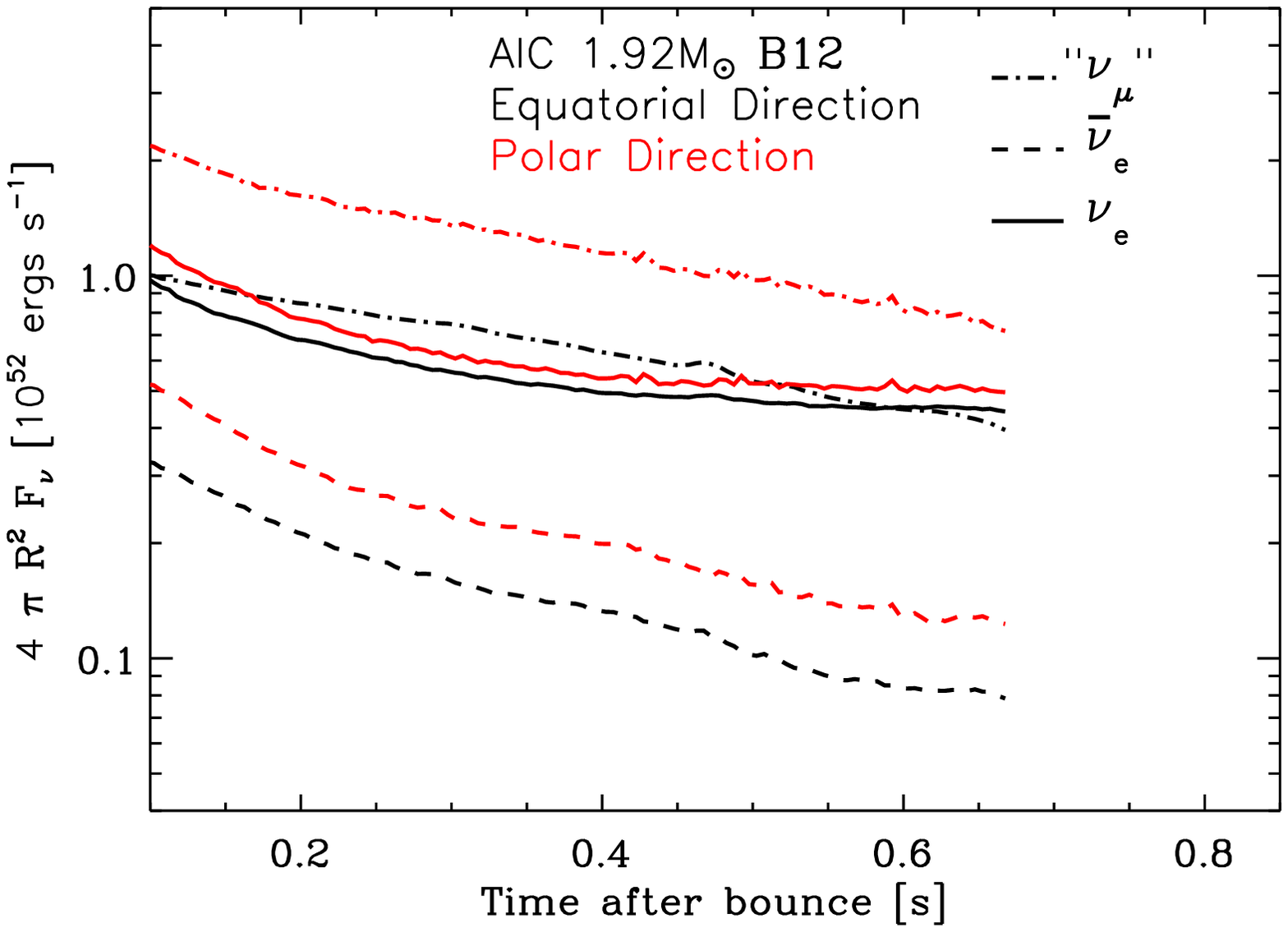}
\caption{
{\it Left:} Time evolution of the electron neutrino (solid line), the anti-electron 
neutrino (dashed line), and the ``$\nu_{\mu}$'' neutrino luminosity (dash-dotted line)
along the equatorial (black) and polar (red) directions, for model B11, 
at $R=$400\,km, and starting 100\,ms after bounce.
Here, neutrino luminosities are integrated over neutrino energy, and correspond to the 
quantity $4 \pi R^2 F_{\nu}(t,R,\theta)$, adopting the flux along the polar direction ($\theta=0$) 
or along the equatorial direction ($\theta=90^{\circ}$).
{\it Right:} Same as left, but for model B12.
Notice the systematic reduction of all neutrino fluxes along the equatorial
direction (neutrinos fluxes decrease in a continuous fashion towards lower 
latitudes), an effect that is essentially geometrical and stems from the larger angle 
subtended by the PNS as seen from higher latitudes. This effect is primarily caused
by the fast rotation of the PNS, rather than its magnetic properties.
Notice also that, for a given time after bounce, the electron neutrino luminosity is reduced 
in the B12 model, and that the anti-electron neutrino and the ``$\nu_{\mu}$''neutrino luminosities
remain comparable to their B11-model counterparts. (See text for discussion.)}
\label{fig_flux}
\end{figure*}

%
%
%
%
%
%
%
%

  The post-bounce neutrino signals of the AIC of the 1.92\,\mo white dwarf we have 
modeled are strongly latitude-dependent\footnote{Note that there was little neutrino field anisotropy 
for the 1.46\,\mo model described in D06, due to the more spherical 
shape of the newly-formed neutron star.} with a systematic increase (irrespective of neutrino type) 
by up to a factor of two from the equatorial to the polar directions (D06).
The gravity-darkened fast-rotating PNS has a much lower temperature along the 
equator, which inhibits the emission, essentially thermal, of ``$\nu_{\mu}$'' neutrinos,
and the flatter density profile along the equator modifies capture rates.
The enhancement in neutrino luminosity at higher latitudes, whatever the species, 
is a purely geometrical effect and results from the increased subtended angle occupied
by the PNS for viewing angles at higher latitudes (see also Walder et al. 2005).
Irrespective of the energy group and the neutrino species, the corresponding 
``radiating'' neutrinosphere is oblate and occupies the largest solid-angle 
for observers along the poles
(see D06 for further discussion on this latitudinal variation of neutrino fluxes).
In addition, we note that, for a given time after bounce, the electron neutrino
luminosity is reduced in the B12 model compared to the B11 model, by $\sim$30\% 
along the equator and $\sim$40\% along the pole, while the luminosities associated with
the $\bar{\nu}_{\rm e}$ and the ``$\nu_{\mu}$'' neutrinos match
their counterparts in the two models within a few percent.
Note that, provided it is not later ejected, the delayed mass accretion, over viscous 
timescales (see \S\ref{sect_GRB}), of the lobes of the progenitor white dwarf has 
the potential to power the neutrino luminosity at a low level over a longer time 
than would typically obtain for isolated neutron stars.

\section{Discussion}
\label{sect_discussion}

  \subsection{Magnetic effects}
\label{sect_MHD}

   The previous section highlighted the main features and phases of the 
AIC of white dwarfs, for both weak and strong magnetic fields. The main effect
of magnetic fields is to trigger a much more powerful explosion, here reaching 
$\sim$1\,B after just 300\,ms after bounce, tapping some free energy of rotation 
of the PNS, and ejecting mass at a rate of 
a few $\times$\,0.1\,\mo\,s$^{-1}$. We thus find that if magnetic pressure becomes comparable
to gas pressure at the neutron star surface, a magnetically-driven wind is initiated whose
properties are boosted when compared to the pure neutrino-driven wind case (i.e., with weak or
negligible magnetic field), the latter with mass loss rates of a few $\times$\,0.001\,\mo\,s$^{-1}$ 
and global ejecta energies of at most $\sim$0.1\,B. Note, in particular, that the free energy of 
rotation is also available at the same level in the weak magnetic-field case (model B11); 
it is only that the means to extract it are insufficient. Magnetically-aided, the explosion is 
sustained at a much higher level, rendering the neutrino component sub-dominant (Fig.~\ref{fig_power}).

    The magnetic contribution to the pressure balance at a few tens of kilometers
can be significant and can deform the PNS compared to the weak field case,
sphericizing the isodensity contours (see, e.g., the bottom panels of Fig.~\ref{fig_entropy}). 
The energy to amplify the magnetic field, to puff-up the neutron star (although mostly along the poles), 
and to drive a more powerful explosion stems from the tapping of core rotational energy 
(where it is the most abundant), which is also manifest in the spin down of the core, 
in analogy with the extraction of angular momentum in accretion disks (Blandford \& Payne 1982).
By $\sim$500\,ms after bounce, the average PNS period\footnote{The average PNS period
is computed assuming solid-body rotation for the same angular momentum budget in the PNS,
with densities greater than 10$^{10}$\,g\,cm$^{-3}$.} 
is 30\% larger in the high magnetic field case, and is $\sim$20\,ms.

  \subsection{Nucleosynthetic yields}
\label{sect_ye}

Guided by the results of Hillebrandt et al. (1984), Wheeler et al. (1998) suggested 
promptly exploding ONeMg cores as the production sites of r-process 
nuclei, given the low electron fraction ($Y_{\rm e} \sim$0.2) and the low entropy 
of their ejecta ($S\sim$\,15\,$k_{\rm B}$\,baryon$^{-1}$; 
see also Qian \& Woosley 1996; Hoffman et al. 1997).
However, radiation-hydrodynamics simulations of the collapse of Fe or ONeMg cores
demonstrated that the explosion can never be prompt (can never be on timescales of a few 
milliseconds) and that, instead, the shock stalls. 
Eventually, in the context of the ONeMg cores of low-mass massive stars or AICs, 
the stalled shock overcomes accretion, because of the extremely steep density 
gradient above the core (Kitaura et al. 2006; D06; Burrows et al. 2007ac).

The results of our present simulations confirm that there is no prompt explosion
in the AIC of white dwarfs and that, indeed, ejecta with low electron fraction
are produced. At low magnetic field, the ejecta are primarily at $Y_{\rm e}$ values between
0.25 and 0.5, but with a total mass of a few $\times$ 0.001\,\mo (see
Fig.~\ref{fig_ye_dist}). At high magnetic fields, the distribution is somewhat altered,
with no material at entropies larger than $\sim$15\,$k_{\rm B}$\,baryon$^{-1}$, and ejecta 
electron fractions extending down to $\sim$0.1. 
More importantly, the total ejecta mass is a factor of $\sim$40 larger 
than in the B11 model and renders such distributions of interest. 
At the end of the B12 simulation, we find a total ejecta mass of 0.15\,\mo, 
with 0.1\,\mo of the material with $Y_{\rm e}$'s between 0.1 and 0.2. and an entropy 
of $S\sim$10\,$k_{\rm B}$\,baryon$^{-1}$. 
Following the argument of Wheeler et al. (1998), in reference to Hoffman et al. (1997),
this indicates the production of nuclei with an atomic mass of up to 190 and motivates
further investigation concerning r-process nucleosynthesis under such conditions.

Woosley \& Baron (1992) and Fryer et al. (1999) used the nucleosynthetic yields
of AICs to estimate their occurrence rate. 
Based on constraints set by a number of isotopic abundances (e.g., $^{62}$Ni, $^{88}$Sr;
see Woosley \& Hoffman 1992), 
Hartmann et al. (1985) suggested that less than 10$^{-5}$\,\mo of material with an 
electron fraction below 0.4 can be ejected per supernova event. 
By the end of the B11 (B12) simulation, a total of $\sim$0.004\,\mo ($\sim$0.15\,\mo) of such 
low-$Y_{\rm e}$ material has been ejected (note that this includes material that has left the grid, 
the maximum computational radius being only 4000\,km), which, together with the isotopic constraints, 
sets the rate to a maximum of $\sim$5$\times$10$^{-5}$\,yr$^{-1}$ for B11-type events, and of 
$\sim$1.3$\times$10$^{-6}$\,yr$^{-1}$ for B12-type events.

Following Woosley \& Baron (1992), an alternate constraint can be imposed by using the solar ratio
of the oxygen to the $^{88}$Sr abundance, assuming oxygen is contributed primarily by core-collapse
explosions of massive-star progenitors, and $^{88}$Sr primarily from AICs.
For the B11 simulation, material with a $Y_{\rm e}$ between 0.4 and 0.45 is ejected 
with a very small cumulative mass of $\sim$5$\times$10$^{-5}$\,\mo at 
800\,ms after bounce, corresponding to a rate constraint of 0.02\,yr$^{-1}$. In the B12 simulation,
more mass is expelled in total, but no ejecta have an electron fraction
above 0.4 after $\sim$200\,ms after bounce, and we indeed find a total ejected mass with an 
electron fraction between 0.4 and 0.45 that is comparable 
to that of the B11 simulation, and a rate of $\sim$0.02\,yr$^{-1}$.
Hence, the second constraint set by the oxygen and $^{88}$Sr solar abundances is not really 
relevant for the AIC we simulate, since we predict nearly all of the nuclei ejected have 
an electron fraction below 0.4.
Overall, if the neutrino- or magnetically-driven winds we witness in the B11 and B12 
simulations persist over a number of seconds, the corresponding galactic AIC rate, as limited 
by nucleosynthetic yields, should be 
at most\footnote{It is unclear how fast the mass loss rate in the B12 model
is going to decrease over the next few seconds, as it must, and, consequently, 
the corresponding cumulative mass loss cannot yet be accurately determined.} 10$^{-6}$\,yr$^{-1}$,
thus from a factor of a few to up to two orders of magnitude lower than that obtained 
in the simulations 
of Woosley \& Baron (1992) or Fryer et al. (1999). Our B12 simulation suggests that if the
progenitor core of such AICs rotates fast and significant magnetic field amplification
occurs after PNS formation, the event rate of the AIC of white dwarfs is doomed to be very small,
of the order of 10$^{-6}$\,yr$^{-1}$, less than four orders of magnitude lower than the 
event rate of massive star explosions.

  We observe a strong pole-to-equator variation of all neutrino fluxes in both the B11 and 
the B12 models. This property was discussed in  D06 and is linked to the fast rotation and oblateness
of the PNS (see also Walder et al. 2005). 
Here, we find that this anisotropy persists with the introduction of magnetic fields.
Moreover, although the anti-electron neutrino and the ``$\nu_{\mu}$''neutrino fluxes
remain comparable in both the B11 and B12 models, we find that the electron neutrino flux 
is systematically lower in the B12 model (for a given direction).

  \subsection{Are AICs $\gamma$-ray-burst/X-ray-flash progenitors?}
\label{sect_GRB}

The AIC of a white dwarf has been proposed repeatedly as a potential GRB progenitor
(Usov 1992,1994; Dar et al. 1992; Yi \& Blackman 1997,1998; Dai \& Lu 1998).
The attractiveness of such collapsed white dwarfs stems partly from 
1) the absence of a progenitor envelope, preventing the choking of possibly relativistic 
ejecta launched at the surface of the newly-formed neutron star, and 
2) an inherently large angular momentum that may be insufficient in massive star 
progenitors to meet the conditions necessary for the collapsar model 
(Woosley 1993; MacFadyen \& Woosley 1999).

Usov (1992, 1994) invoked the generation of a relativistic electron-positron
plasma from such a fast-rotating highly-magnetized neutron star, powered over 
a timescale of seconds, leading to the emission of X-rays and $\gamma$-rays 
at the wind photosphere. A similar model was proposed by Yi \& Blackman (1997,1998) 
and Dai \& Lu (1998), in which a magnetized millisecond pulsar powers a fireball 
by magnetic dipole radiation.
Alternatively, Dar et al. (1992) suggested that electron-positron pair production from
neutrino-antineutrino annihilation outside the PNS would trigger
a GRB, provided the neutron star environment is not polluted with more than 3$\times$10$^{-4}$\,\mo
of baryons. 

  In the context of standard core-collapse supernova explosions, neutrino-antineutrino 
annihilation does not lead to appreciable net energy gain for the wind or ejecta because the
energy tends to be deposited deep in the PNS, or at best in the cooling region 
in the vicinity of the neutrinosphere,
thereby contributing negligibly to the supernova explosion (Cooperstein et al. 1987). 
Additionally, the annihilation cross section is strongly enhanced for large-angle 
collisions, favoring non-spherical configurations like disks or tori (Janka 1991; 
Ruffert \& Janka 1997). The disk-like morphology of the neutron stars formed
in the AIC of white dwarfs, and the additional pinching of isodensity contours
in the polar regions, seems to offer an interesting prospect. 
Following the method of Ruffert \& Janka (1997) and using the final
configuration of the B12 model we estimate an
$\nu_{\rm e}$\,$\bar{\nu}_{\rm e}$ annihilation power of 0.01\,B\,s$^{-1}$, 
deposited primarily along the polar direction, but within 20--50\,km of the PNS center.
This is a rough estimate, but it suggests that if our B11/B12 models are sensible
representations of the AIC of white dwarfs, then little can be gained from
neutrino-antineutrino annihilation, in particular in comparison with the power 
injected by magnetic means. This is the case at a few hundred milliseconds after bounce,
but it may change at later times of many seconds.

In the general context of core-collapse supernovae, Tan et al. (2001) investigated 
the acceleration to relativistic speeds of the surface layers of the progenitor 
as the shock, generated at bounce, breaks out. The AIC of white dwarfs presents an interesting 
context for such an effect, given the absence of a progenitor envelope.
But the explosion energies and the ejecta masses reported in the spherical simulations 
of the AIC of white dwarfs (Woosley \& Baron 1992; Fryer et al. 1999) were low and
Tan et al. confirmed the conclusion of those authors that the AIC of white dwarfs cannot 
lead to any sizable $\gamma$-ray emission or relativistic ejecta.
In D06, and reproduced here as well for the B11 model, the explosion
is strongly underenergetic for weak magnetic fields (reaching only $\sim$0.1\,B after about
500\,ms, with a power that has stabilized to $\sim$0.1\,B\,s$^{-1}$ and that may be sustained
for a few seconds).
Most of the power is in the form of a neutrino-driven wind, 
with an associated mass loss rate of a few $\times$ 0.001\,\mo\,s$^{-1}$.
The original blast that breaks out of the polar caps of the progenitor white dwarf would
represent the best option in the context of the Tan et al. study, but the shock, debilitated by
neutrino losses and the photodissociation of nuclei, stalls only mildly. It is not re-energized
by neutrino energy deposition. Rather, it breaks out of the white dwarf by virtue of the abrupt
density gradient. The resulting early blast amounts to a mere 10$^{-4}$\,B, and despite
the low baryon loading of the ejecta ($\sim$10$^{-4}$\,\mo), they reach velocities
of only a tenth of the speed of light\footnote{Note here that the low-density ambient medium surrounding 
the progenitor white dwarf may be slightly inhibiting further acceleration, 
but Fig.~11 of Tan et al. 2001 suggests that this shock breakout is in any case 
symptomatically underenergetic.}.
In model B12, the explosion energy goes above 1\,B, but the wind mass loss rate
is commensurate with this energy gain, and is on the order of 0.1\,\mo\,s$^{-1}$.
As a result, the production
of a GRB via the Tan et al. scenario is again not viable. The asphericity of the explosion
does not dramatically alter this conclusion, since, while the explosion energy is deposited
in a smaller volume, the mass flux is also confined to a smaller opening angle and leads to
higher effective baryon loading.
The spherically-symmetric, adiabatic, steady-state, LTE model of a super-Eddington 
wind by Paczynski (1990) supports the same conclusion. Given a rate of energy deposition 
of a 0.1--1\,B\,s$^{-1}$ and a mass loss rate between 
a few $\times$ 0.001 and $\sim$0.7\,\mo\,s$^{-1}$ for the B11 and B12 models, respectively, the 
temperature at the photosphere would be vanishingly small in both cases, 
with a photon luminosity of at most 10$^{40}$\,erg\,s$^{-1}$.
The baryon loading of the ejecta that occurs in the B12 model, associated with the 
magnetic-driving of the wind, nulls the benefit of the additional explosion power.
The low photospheric temperature obtained here suggests that, since the
ejecta become optically thin at such a late time, no radiation would
be visible in the optical. The negligible amount of ejecta material 
with a $Y_{\rm e}$ of 0.5, which corresponds to radioactive $^{56}$Ni, means that
there is no energy source to counteract the strong adiabatic cooling of the neutrino-
or magnetically-driven wind. It seems that neutrinos and gravitational waves would be 
the only significant signatures of the explosion.

The AIC of a white dwarf might give birth to a magnetar, following magnetic-field 
compression and amplification in the fast rotating neutron star. 
Yi \& Blackman (1997,1998), and Dai \& Lu (1998) suggested that such magnetars could, through extraction
of rotational energy, power Poynting-flux dominated ejecta and lead to GRBs at cosmological
distances. Indeed, during the cooling phase of such fast-spinning highly-magnetized
neutron stars, the millisecond proto-magnetar wind may transition from non-relativistic 
and thermally-driven to magneto-centrifugally-driven, and finally to relativistic and 
Poynting-flux dominated (Thompson et al. 2004; Bucciantini et al. 2006; Metzger et al. 2006; 
Thompson 2007). This scenario has in fact the potential to make 
the AIC of white dwarfs detectable in the $\gamma$-ray range.
Furthermore, based on the local Universe distribution of short-hard GRBs among all galaxy types, rather
than exclusively with young stellar populations, Chapman et al. (2006) have proposed 
the AIC of white dwarfs as an alternative to massive star progenitors for the formation of magnetars,
whose soft $\gamma$-ray emission is associated with giant-flaring events.
We concur in principle with this proposition, although the AIC rate of at most 
$\sim$10$^{-6}$\,yr$^{-1}$ for B12-type events that we infer in \S\ref{sect_ye} suggests
that magnetars originating from the AIC of a fast-rotating white dwarf 
should be significantly under-represented in the magnetar population.

An AIC analogue to the collapsar model (Woosley 1993; MacFadyen \& Woosley 1999) is an 
obvious alternative, with both the transition of the PNS to a black hole and energy deposition 
in the excavated polar regions at a high rate of a few $\times$ 0.1\,B\,s$^{-1}$.
In the context of the AIC of white dwarfs, the potential for formation of a black hole seems uncertain,
as it might be difficult in nature to ultimately meet the basic mass requirement of a 
2--2.5\,\mo (baryonic) neutron star for black-hole formation 
(Baym et al. 1971; Arnett \& Bowers 1977). In D06, we studied two models,
with 1.46\,\mo and 1.92\,\mo masses. The PNS mass was larger in the lower mass 
progenitor because in the 1.92\,\mo model much of the mass ended in a massive disk.
Higher-mass AIC progenitors would need even more angular momentum to achieve a stable
equilibrium configuration, which would likely result in the formation of even higher-mass disks
on extended quasi-Keplerian orbits. At the same time, magnetic stresses at the surface of the
neutron star lead to such high mass loss rates that the PNS mass paradoxically decreases with time 
(the PNS mass in the B12 model decreases to $\sim$1.3\,\mo at $\sim$500\,ms after bounce).
It, thus, seems that accretion, leading to black-hole formation, would have to operate
over a very extended time, and it is not clear that given the explosion power that can be 
obtained (witness the results for the B12 model) whether accretion will ever win against explosion.
A further obstacle to high accretion rates in the AICs is the extended configuration of the 
mass distribution outside the neutron star, with progenitor lobes located 
at $\sim$1000\,km, rather than around $\sim$100\,km in the ideal setup for the collapsar model
(Woosley 1993; MacFadyen \& Woosley 1999).
The outcome here is closely related to the original mass distribution of the white dwarf
at the time of collapse.
Current SPH simulations of coalescing white dwarfs, one formation channel of AICs,
suggest that a significant fraction of that material goes into a Keplerian disk, 
loosely-bound and, thus, not directly involved in the collapse/bounce of the core, 
nor with the subsequent explosion.
This material may represent a substantial amount of mass, perhaps up to 1\,\mo, but is located
at least at 1000\,km away from the neutron star or would-be black hole, so that the actual
accretion rate is very small. In the configuration of the 1.92\,\mo model described here,
we have 0.5\,\mo available in the disk, with an associated viscous accretion timescale of $\sim$100\,s 
(using an $\alpha$ parameter of 0.1; Shakura \& Sunyaev 1973), and hence,
a feeble accretion rate of 5$\times$10$^{-3}$\,\mo\,s$^{-1}$. This is a factor
of a hundred lower than the minimum value quoted in the context of the collapsar
model. This extended configuration for AICs is in fact reminiscent of
the fallback scenario of Rosswog (2007) for double neutron star mergers, wherein
material is placed on eccentric orbits during the merger event. Note that, here again,
accretion onto a black hole is a prerequisite for the powering of a relativistic jet,
and, ultimately, $\gamma$-ray radiation.
In the context of AIC, such accretion may simply extend the PNS-wind phase 
over a longer period, and not lead to the formation of a black hole.

  \subsection{Ultimate fate of the AIC of white dwarfs}

The AIC of white dwarfs seems very unfavorable as a progenitor of $\gamma$-ray, or even optical
bursts, and direct detection of such events might instead come from neutrino
or gravitational wave detectors (see \S\ref{sect_sim} and \S10 of D06). 
Indirectly, we may infer their existence through the r-process nuclei ejected during the 
explosion (\S\ref{sect_ye}).

Another avenue is to search for them in the neutron star population.
Fallback following SN explosions of massive progenitors is expected
to bury the PNS and weaken its surface magnetic field\footnote{
In this context, the mass accreted after a delay starts with a
very weak initial field. Little compression ensues and the spin up is moderate,
so that, despite the presence of velocity shear in the region of accretion,
we expect only modest field amplification.}, as in magnetic 
neutron stars accreting from a companion star
(see, e.g., Geppert \& Urpin 1994; Payne \& Melatos 2004,2007).
This argument would suggest that magnetic neutron stars may originate
from the AIC of white dwarfs, since these are exempt from such fallback.
Observations of magnetic neutron stars in binary systems have indeed been
interpreted in this way (see, e.g., Taam \& van den Heuvel 1986, Ergma 1993;
van den Heuvel \& Bitzaraki 1995; van Paradijs et al. 1997), although our simulations
suggest that some ``fallback'', albeit modest, could occur from the progenitor lobes of
the white dwarf and bury the large surface fields of $\sim$10$^{15}$G that we see 
in the B12 model a few hundred milliseconds after bounce. 
Hence, in this context, whether additional material from the progenitor lobes of the white dwarf
accretes onto the highly-magnetized PNS may determine whether it
will retain its magnetar-like fields or transition to a weaker-field pulsar.
Despite the extraction of angular momentum by magnetic fields, the PNS
remains a millisecond-period rotator. Further accretion from the Keplerian
disk would mitigate the loss of angular momentum, and, thus, AICs of white dwarfs
should consistently lead to rapidly-rotating pulsars after this original phase.
On longer timescales of hours or days, however, magnetic torquing may be more 
in evidence and may spin down the PNS as it transitions to a magnetar-like object.
In this case, the remnant could be a highly-magnetized object, with a much reduced
rotation rate, a correlation that has been found in the pulsar population 
(Vranevsevic et al. 2007).
Interestingly, if such accretion is not significant, this material could be found 
in a debris disk, as recently observed around an isolated neutron star (Wang et al. 2006).

   Finally, we reported in D06 that in the simulation for the
1.46\,\mo AIC model of Yoon \& Langer (2005), carried over 180$^{\circ}$ (i.e. including
both hemispheres), perfect north-south symmetry was preserved. We have used this observation
to carry out all subsequent investigations in a 90$^{\circ}$ quadrant, thus in just one hemisphere.
We expect that the AIC of white dwarfs leads to essentially perfect north-south symmetry,
a property that applies more generally to the explosion following the collapse
of ONeMg cores, universally characterized by a steep density profile above the core.
Consequently, these explosions should involve no significant
kick to the PNS. This supports the notion that the AIC of white dwarfs
can, at least theoretically, contribute to the neutron star population of 
globular clusters (see, e.g., Bailyn \& Grindlay 1990; Ivanova et al. 2006).
But given the very low inferred rates of AICs (in particular for the B12-like events), 
the ONeMg cores of low-mass 
massive stars are more likely candidates. Similarly, this also supports the
association of collapsed ONeMg-cores with the populations of low-velocity, 
low-mass neutron stars found in low-eccentricity double neutron star 
systems (see, e.g., van den Heuvel 2007).

\section{Conclusion}
\label{sect_conclusion}

  We have presented multi-group, flux-limited diffusion, magnetohydrodynamics simulations 
of the Accretion Induced Collapse (AIC) of white dwarfs, focusing on the 1.92\,\mo model
of Yoon \& Langer (2005). The main incentives were to gauge the relevance and estimate the 
magnitude of MHD effects
in such collapse events, and thereby to address a potential limitation of the 
simulations in D06. The key ansatz in the present work is that, although 
the magnetic field distribution and magnitude in the progenitor white dwarf are
unknown, fast and differential rotation in the collapsed object is expected to lead 
to an exponential amplification of any initially weak poloidal field through the magnetorotational
instability, reaching large values at saturation.
In this context, the magnitude of the post-bounce MHD effects we report from our simulations
will thus be determined, in a more physical sense, by the angular momentum 
budget rather than by the initial field in the pre-collapsed white dwarf.
Provided angular momentum is abundant, we find that MHD effects lead to
large quantitative changes when compared with an equivalent situation ignoring magnetic fields.
Said differently and in a broader sense, we expect significant MHD effects in core-collapse
events with fast rotation, effects that cannot be ignored without compromising the
conclusions \footnote{Non-axisymmetric instabilities of such fast-rotating white dwarfs
may alter the angular momentum distribution after bounce (Ott et al. 2006), 
with the potential to modify magnetic-field amplification rate as well as the accretion 
rate we observe in our simulations. Numerical simulations of the kind presented here 
but in 3D are needed to quantify these effects.}.

  Perhaps the most striking property of the AIC of a 1.92\,\mo white dwarf model
with large initial magnetic fields is that the explosion energy at just a few
hundred milliseconds after bounce is no longer an anemic 0.01\,B, but is on the order
of one Bethe, with an ejecta mass raised from a few $\times$ 0.001\,\mo to $\sim$0.1\,\mo.
Significant rotational energy is extracted from the core with the inclusion of magnetic fields, 
directly seen in the decrease in rotational energy and the spin down of the PNS.
This energy is transformed into magnetic energy that powers the explosion.
For higher B-fields, the density distribution at the surface of the neutron 
star is more spherical, the 10$^{10}$\,g\,cm$^{-3}$ and  10$^{11}$\,g\,cm$^{-3}$ 
density contours stretching further out along the pole, aided by magnetic stresses.
We find that mass loading of the ejecta is so severe that no relativistic 
high-energy electromagnetic transient
can be expected during the PNS phase. In the past, the AIC of white dwarfs
has been invoked as a potential GRB progenitor, but this was usually done in ignorance
of the neutrino-driven wind that prevails during the first tens of seconds after the formation
of the PNS. MHD effects seem to make this situation even worse, not better,
by loading magnetically-driven ejecta with 100 times more baryons.

In the single-degenerate formation scenario, magnetic torques operating over evolutionary 
timescales might force the white dwarf into solid-body (and, perhaps, critical) 
rotation, and prevent its growth much beyond the canonical Chandrasekhar mass, thus preventing a 
transition to a black hole after PNS formation. However, in the double-degenerate scenario,
differential and fast rotation results after the merging event and a super-Chandrasekhar white 
dwarf may form. Ultimately, transition to a black hole may occur if the total PNS mass 
exceeds the general-relativistic limit for gravitational stability.
The transition to a black hole, by shutting off the wind from the PNS,
may then lead to a $\gamma$-ray burst or an X-ray flash, but the 2--2.5\,\mo mass limit
for black hole formation restricts the number of adequate AIC progenitors to a very small number,
difficult to estimate, but perhaps to one in a hundred AICs.
The coalescence of two white dwarfs\footnote{It even seems that these two
white dwarfs would have to have comparable masses. Otherwise, the secondary would likely get 
disrupted and with its matter dispersed in a disk around the primary, rather than merging with 
the primary (Guerrero et al. 2004).} 
seems the more favorable configuration, but no more than a few $\times$ 0.1\,\mo of material placed at 
large distances (few 1000\,km) would be available for accretion, and this over a viscous timescale
of $\sim$100\,s and, thus, with a feeble accretion rate of at most 5$\times$10$^{-3}$\,\mo\,s$^{-1}$. 
In the unlikely event that all these necessary conditions were met, the relativistic 
beaming (with opening angle $\theta_{\rm j}$) of the radiating ejecta would decrease 
the probability of $\gamma$-ray detection by a factor $0.5 (1 -\cos\theta_{\rm j})$,
i.e., 2--3 orders of magnitude for a 5$^{\circ}$--15$^{\circ}$ opening-angle
\footnote{This relativistic beaming of $\gamma$-ray radiation is inferred from
the achromatic steepening breaks of some GRB optical afterglow lightcurves 
(Kulkarni et al. 1999; Harrison et al. 1999).}.  
Together with the constraints from nucleosynthetic yields of B12-like events,
the 1 in a 100 AICs that may reach the black-hole mass limit, and the inherent beaming of
the resulting $\gamma$-ray radiation, we estimate the detection rate of $\gamma$-ray bursts associated 
with AICs to be on the order of 10$^{-10}$\,yr$^{-1}$,
significantly lower than the average rate per galaxy of 4 $\times$10$^{-7}$\,yr$^{-1}$ 
inferred from GRBs detected with BATSE (Zhang \& M{\'e}sz{\'a}ros 2004).

Moreover, the absence of nickel in the ejecta precludes any significant optical signature
for the event, which should thus remain obscure at all wavelengths. If the AIC occurs in 
our Galaxy, possible signatures should instead be from the neutrino burst that accompanies 
the formation and the cooling of the PNS, and gravitational waves from the very aspherical 
bounce and the anisotropic neutrino emission (Ott 2007).

The AIC of white dwarfs may, however, have a significant impact on the interstellar medium
composition in r-process nuclei. We find that the AIC of rapidly-rotating 
white dwarfs leads to explosions
that eject copious amounts of neutron-rich material, with an electron fraction 
reaching down to $\sim$0.1--0.2, depending on the magnitude of the neutrino/anti-neutrino
absorption by the ejecta. If not magnetically-aided, the neutrino-driven wind leads, however,
to a low ejecta mass of a few $\times$ 0.001\,\mo, which, given the low occurrence rates for AIC,
makes these yields somewhat irrelevant. If magnetically-aided, as in the simulation B12 presented
here, about 0.1\,\mo of material is ejected with an electron fraction between 0.1 and 0.2,
making the AIC of white dwarfs a possible r-process site, as promoted in an analogous context 
by Wheeler et al. (1998). More importantly, constraints placed by solar isotopic abundances
require that such pollution of the interstellar medium by magnetically-driven winds of AICs
occurs with an event rate of at most $\sim$10$^{-6}$\,yr$^{-1}$. If we believe the
context adopted for this work, this suggests that the circumstances leading to such fast 
rotating white dwarfs are rare.

An important ramification of this work is that, free of significant fallback, 
the AIC of white dwarfs offers a possible scenario for the formation of highly magnetized 
neutron stars, potentially with magnetar-like surface fields of 10$^{15}$\,G, with
magnetically-torqued cores that spin down over many seconds. Here again, though, the low
event rate of such AICs should result in their contributing only modestly to the magnetar 
population.

\acknowledgments

We thank Jim Liebert, Jeremiah Murphy, Martin Pessah, Todd Thompson, and Stan Woosley
for fruitful discussions and their insight. We acknowledge support for this work
from the Scientific Discovery through Advanced Computing
(SciDAC) program of the DOE, under grant numbers DE-FC02-01ER41184
and DE-FC02-06ER41452, and from the NSF under grant number AST-0504947.
E.L. thanks the Israel Science Foundation
for support under grant \# 805/04, and C.D.O. thanks
the Joint Institute for Nuclear Astrophysics (JINA) for support under
NSF grant PHY0216783. This research used resources of the National
Energy Research Scientific Computing Center, which is supported by the
Office of Science of the U.S. Department of Energy under Contract No. 
DE-AC03-76SF00098.

\end{document}